\documentclass[sigconf]{acmart}

\usepackage{multirow}
\usepackage[T1]{fontenc}
\usepackage{caption}
\usepackage{subcaption}
\usepackage{multicol}

\usepackage{tabularx}
\usepackage{graphicx}
\usepackage{adjustbox}
\usepackage[english]{babel}

\usepackage{xcolor, colortbl}
\usepackage{xspace}
\usepackage[flushleft]{threeparttable}
\usepackage{rotating}
\usepackage{fancybox}
\usepackage{color}
\usepackage{balance}
\usepackage{enumitem,kantlipsum}

\definecolor{grayboxcolor}{HTML}{f2f2f2}
\newcommand{\conclusion}[1]{%
	\begin{center}\noindent\thicklines\setlength{\fboxsep}{3pt}\fcolorbox{black}{grayboxcolor}{\begin{minipage}{3.4in}\textit{\textbf{#1}}\end{minipage}}\end{center}}

\newcommand{\rqiisaner}{\textcolor{black}{RQ1: Can the augmentation approach improve the NLU's performance?}}
\newcommand{\rqiiisaner}{\textcolor{black}{RQ2: Does the augmentation approach increase the NLU's confidence in its classification?}}

\title{A Transformer-based Approach for Augmenting Software Engineering Chatbots Datasets}

\author{Ahmad Abdellatif}
\affiliation{%
  \institution{University of Calgary}
  \city{Calgary}
  \country{Canada}}
\email{ahmad.abdellatif@ucalgary.ca}

\author{Khaled Badran}
\affiliation{%
  \institution{Concordia University}
  \city{Montreal}
  \country{Canada}}
\email{k\_badran@encs.concordia.ca}

\author{Diego Elias Costa}
\affiliation{%
  \institution{Concordia University}
  \city{Montreal}
  \country{Canada}}
\email{diego.costa@concordia.ca}

\author{Emad Shihab}
\affiliation{%
  \institution{Concordia University}
  \city{Montreal}
  \country{Canada}}
\email{emad.shihab@concordia.ca}

\begin{document}

\begin{abstract}

Background: The adoption of chatbots into software development tasks has become increasingly popular among practitioners, driven by the advantages of cost reduction and acceleration of the software development process. Chatbots understand users' queries through the Natural Language Understanding component (NLU). To yield reasonable performance, NLUs have to be trained with extensive, high-quality datasets, that express a multitude of ways users may interact with chatbots. However, previous studies show that creating a high-quality training dataset for software engineering chatbots is expensive in terms of both resources and time.
Aims: Therefore, in this paper, we present an automated transformer-based approach to augment software engineering chatbot datasets. Method: Our approach combines traditional natural language processing techniques with the BART transformer to augment a dataset by generating queries through synonym replacement and paraphrasing. We evaluate the impact of using the augmentation approach on the Rasa NLU's performance using three software engineering datasets. Results: Overall, the augmentation approach shows promising results in improving the Rasa's performance, augmenting queries with varying sentence structures while preserving their original semantics. Furthermore, it increases Rasa's confidence in its intent classification for the correctly classified intents. Conclusions: We believe that our study helps practitioners improve the performance of their chatbots and guides future research to propose augmentation techniques for SE chatbots. 
\end{abstract}

\maketitle

\section{Introduction}
Chatbots have proven themselves to be a game changer in a variety of domains from personal assistant to customer services. With their benefits in saving time and cost, chatbots have made significant advances in various fields~\cite{Storey2016FSE}.
The increased popularity and proven benefits of chatbots are driving software engineering (SE) practitioners to develop chatbots to help developers in various SE tasks. For example, \citet{Lin_2020BotSE} developed the MSABot, a chatbot that assists developers in building and managing microservices projects (e.g., setting microservices project parameters). \citet{Abdellatif2020EMSE} developed the MSRBot to answer questions related to software projects (e.g., ``Who fixed bug 5?'').

Through natural language, chatbots enable users to communicate with different services intuitively. To understand users' queries (i.e., messages), chatbots leverage a Natural Language Understanding (NLU) component~\cite{Abdellatif2020EMSE,Lin_2020BotSE,Abdellatif_2021TSE}. In essence, NLUs use AI and natural language processing techniques to extract structured information (the intent of the user's query and related entities) from unstructured input text. To use NLUs effectively, chatbot developers need to obtain or craft high-quality datasets containing a variety of user queries to train the NLU in extracting the intention behind the user's questions. Prior work shows that the performance of the NLU is directly related to the quality and diversity of the dataset used in its training~\cite{Abdellatif_2021TSE}. Indeed, including syntactically diverse queries with the same semantics in their training datasets to train the NLU in the different ways users may ask for the same information.~\cite{sentenceStructureRasa_link,sentenceStructureLUIS_link}. For example, the queries ``List the developers who resolved issue 5'', ``Who fixed bug 5?'', and ``Which developer fixed issue 5?'' have the same semantics (identify the developer who fixed a specific bug) but different sentence structures.

Crafting a diverse and high-quality dataset is one of the most costly and time-consuming tasks in chatbot development~\cite{Dominic20BotSE,Abdellatif2020EMSE,Abdellatif_MSR2020}. Chatbot developers need to brainstorm a variety of training queries in order to familiarize the NLU with new terms (synonyms replacement) and diverse sentence structures (paraphrasing)~\cite{Utteranc5:online,Abdellatif2020EMSE}. Previous studies show that the lack of high-quality datasets is a limiting factor for the efficiency of chatbots~\cite{Dominic20BotSE,Abdellatif2020EMSE}.
For example, \citet{Dominic20BotSE} reported that the absence of training queries limited their chatbot performance. Likewise, \citet{Abdellatif2020EMSE} stated that the MSRBot failed to classify some user queries correctly because of the scarcity of training data. Consequently, this data problem hinders the practitioners' ability to develop more efficient SE chatbots, as the training dataset would need to be crafted and augmented manually. Moreover, there is a number of posts on Stack Overflow where chatbot developers ask for more data to enhance the NLUs' performance~\cite{lackDataSOF_link1,lackDataSOF_link3}.

When practitioners create the initial set of training queries, \textit{Augmentation} techniques are often used to create and incorporate new training queries into the dataset~\cite{Abdellatif_MSR2020}. 
This process is key for many machine learning applications where data scarcity is a major limiting factor for model's performance. 
A number of studies have focused on evaluating different augmentation approaches to improve different machine learning applications in the field of sentiment analysis~\cite{Marivate_2020MLKE,Imran2022ASE} and hate-speech detection~\cite{Rizos_2019CIKM}. 

Inspired by recent breakthroughs in language models for different SE tasks~\cite{Ciniselli2022TSE,Tufano2022ICSE,Ciniselli2022MSR}, we explore a transformer-based augmentation approach for SE chatbots that emulates the way in which chatbot developers augment their datasets~\cite{Utteranc5:online,Abdellatif2020EMSE}. More specifically, the augmentation approach takes as input a few training queries and uses them to augment more queries by replacing some words with their synonyms (synonyms replacement). Then, it uses a fine-tuned BART transformer to change the query structure (paraphrasing). To evaluate the performance of the augmentation approach, we perform an empirical study by applying the Augmentation Approach to three well-crafted datasets that represent distinct SE-related tasks, namely 1) Repository: questions exploring software project data, 2) Ask Ubuntu: technical questions from the Ubuntu Q\&A community on Stack Exchange, and 3) Stack Overflow: questions commonly asked by developers on Q\&A websites. 
These datasets include a total of 767 queries covering 19 different intents (e.g., LookingForCodeSample). To put the Augmentation Approach results into perspective, we compare the Rasa NLU's performance without augmenting any query to the training dataset (Baseline), and augmenting human queries to the training dataset (Human). Our study is formalized through the following research questions:\\
\textbf{\rqiisaner}
Overall, the Augmentation Approach shows promising results in improving the Rasa's performance, where it marginally increases the Rasa's performance by up to 3.2\% compared to Baseline. Furthermore, the augmented queries have different sentence structures with the same semantics as the input queries. In cases where there was no improvement, the Approach augments queries with small modifications, which could result in overfitting the Rasa. On the other hand, the Human augmented queries improve Rasa's performance across all datasets compared to Augmentation Approach.
\\
\textbf{\rqiiisaner}
Training Rasa using the approach augmented queries increases the Rasa's confidence in its intent classification for the correctly classified intents. In some cases, the Augmentation Approach outperforms Humans in terms of confidence scores for correctly classified intents. For the misclassified intents, the results show that the augmented queries in the Human and Augmentation Approach experiments increase Rasa's confidence in the misclassified intents.
\\

\noindent Our study makes the following contributions:
\begin{itemize}
\item We propose an approach that uses synonym replacement and paraphrasing techniques to augment queries for chatbots in the software engineering domain. 

\item We explore the impact of using the augmentation approach on the NLU's performance using three datasets from the SE domain and different use cases that vary in the number of initial training queries.

\item We provide a replication package containing the implementation of the augmentation approach as a prototype tool and our results~\cite{Augmenti93:online} to facilitate the replication and accelerate future research in the area.

\end{itemize}

\section{Background}
\label{sec:background6}
Chatbots are software bots that interact with users through natural language~\cite{Lebeuf2019BotSE}. This simple method of interaction is what gives chatbots their appeal and makes them a suitable conduit between users and services, such as in customer service~\cite{Lebeuf2018IEEE}. To facilitate this chat-like interaction, modern chatbots leverage the natural language understanding (NLU) component, which extracts structured information from unstructured text (user's query). Typically, the NLU component extracts two key pieces of information from the user's query; the intent and entities. The \textbf{intent} represents the intention/goal behind the user's query, while \textbf{entities} are important keywords in the query. For example, when a user asks ``\textit{What are the fixing commits for bug 5391?}'', the intent is to know which commits fixed a specific bug in the project (`FixingCommit' intent), whereas the bug number (`5391') is the entity. When interacting with a chatbot, users are free to express the same intent in different ways. For example, the queries ``Show the fixing commits for issue 5391'' and ``What changes solved 5391?'' have the same intent (`FixingCommit') but different syntax.

It is critical for chatbots to have a robust NLU that extracts the intent from the users' queries correctly as it gives the chatbot an accurate assessment of the users' intentions, leading it to take the right course of action (send a reply or perform a task). 
In contrast, a poorly performing NLU that misclassifies the users' intent will lead the chatbot to reply incorrectly and/or perform a wrong action which has a direct and negative impact on the satisfaction of the chatbot users~\cite{Lebeuf2018IEEESoftware}.

When the NLU extracts an intent, it also returns a confidence score corresponding to that intent. The confidence score shows how confident the NLU is in its intent classification, and it has a value that ranges between 0 (i.e., not confident) to 1 (i.e., fully confident). Chatbot developers use the confidence score to determine whether the chatbot has understood the user's query well enough (high confidence score), in which case, the chatbot should perform an action. Otherwise, if the user's query is not clear enough (low confidence score), the chatbot asks the user to clarify the query in order for the chatbot to better understand the intent~\cite{Abdellatif_2021TSE}.

Chatbot developers brainstorm to come up with training queries at the early stages of the chatbot development cycle to train the NLU on different ways a user could ask about specific intent~\cite{Abdellatif2020EMSE}. Then, they augment more training queries by replacing some words with their synonyms and re-writing (paraphrasing) the original training queries in different ways~\cite{Abdellatif_MSR2020}. However, augmenting the chatbot training queries is a resource and time consuming task~\cite{Dominic20BotSE,Abdellatif2020EMSE}.

In our study, we want to investigate the impact of an augmentation approach on the NLU's performance in terms of intents classification and confidence score. Intent classification with high confidence is critical to ensure that chatbots correctly understand and answer the user's question, which improves the user experience with the chatbot.

\section{Approach}
\label{sec:approach6}
\begin{figure}
	\centering
 	\includegraphics[width=\linewidth]{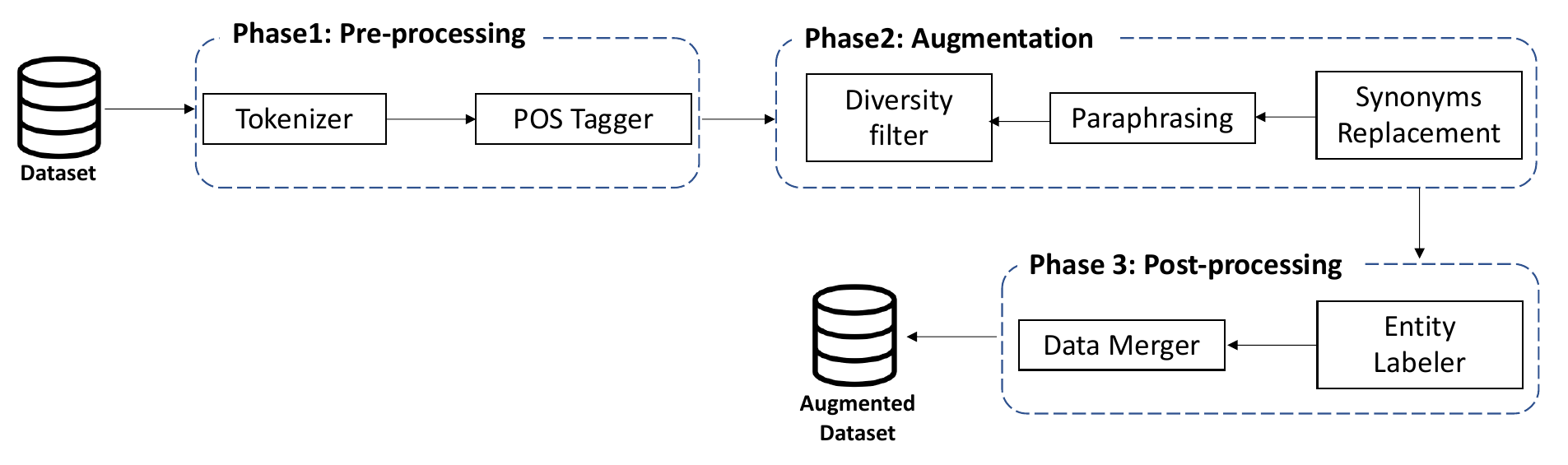}	
\vspace{-0.15in}
	\caption{An overview of the augmentation approach.}
	\label{fig:approach-overview}	
  \vspace{-0.15in}
\end{figure}

\begin{figure}[t!]
	\centering
	\includegraphics[width=\linewidth]{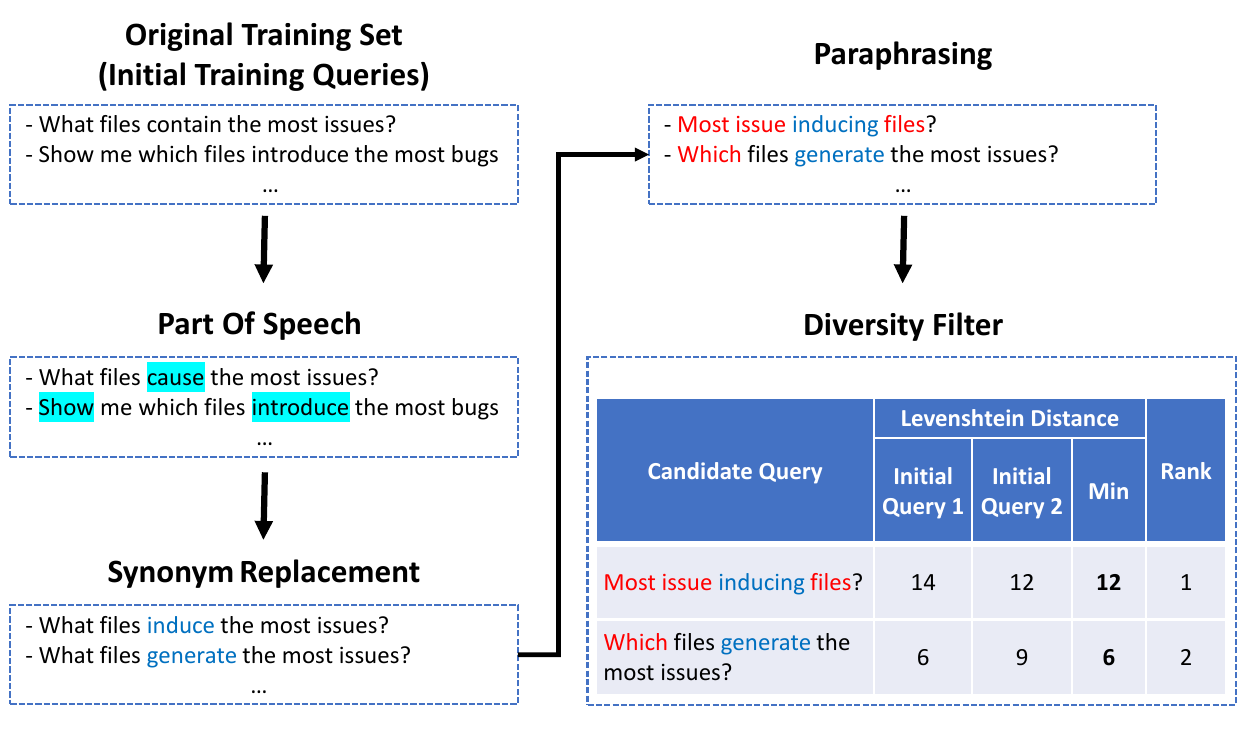}	
\vspace{-0.2in}
	\caption{A working example of the augmentation approach.}
	\label{fig:working_example}	
  \vspace{-0.2in}
\end{figure}

The key idea of the augmentation process is that by using a small initial set of training queries (called the original training set) as input, we can generate new queries that retain the same semantics while having different terms/keywords and brand new sentence structures. Figure~\ref{fig:approach-overview} shows the components of the augmentation approach. Initially, the augmentation approach takes a query from the original training set as an input, then it tokenizes the query and extracts the part-of-speech for each token in the query. Next, the augmentation approach generates new queries (candidate queries) by introducing new keywords and paraphrasing the input query. Then, it filters the candidate queries and keep only the queries with the highest potential of improving the NLU's performance. Finally, the approach labels the entities in the selected queries and merges them with the original training set to obtain the final (augmented) training set. 
In this section, we detail each component of the augmentation approach. Also, we showcase an end-to-end example (Figure~\ref{fig:working_example}) to demonstrate how each component works.

\underline{\textit{Tokenizer:}} 
Commonly used in an NLP pipeline, we start our augmentation approach by tokenizing the input data, to better process and augment text, such as identifying part of speech and replacing synonyms. Thus, in this component, we split each input query in the original training set into tokens using a pre-trained model from the SpaCy library.

\underline{\textit{Part-of-Speech (POS) Tagger:}} This component identifies the POS (e.g., verb, noun, adjective) for each token in the query. This makes the augmentation approach more flexible as it can apply synonyms replacement on specific POS tokens. For example, in case the dataset already has diverse synonyms for noun tokens, then the chatbot developers may want to diversify the dataset by having more synonyms of verbs. The working example in Figure \ref{fig:working_example} showcases the POS tagging component identifying and labelling the verb tokens (i.e., cause, show, and introduce) in the input queries in the original training set.

\underline{\textit{Synonyms Replacement:}} 
Given the tagged tokens from the POS Tagger component, the Synonyms Replacement component replaces certain tokens (e.g., verbs, nouns) with their synonyms to obtain new candidate queries. The goal here is to familiarize the NLU with a large variety of similar terms that might appear in the users' queries, but have not been exposed to the NLU during its training. To obtain the list of synonyms for a token, one could use any of the available thesauruses such as WordNet~\cite{wordnet_online} and PyDictionary~\cite{geekprad94:online}. However, those are general purposes thesauruses and not tailored for SE specific terminologies. For example, when looking for synonyms to the term `bug', the WordNet thesauruses returns `germ', `microbe', and `hemipteron'. Since our goal is to augment the SE chatbot training dataset, we opted to use a thesaurus that is specialized for SE to capture the specific language and terminologies used in the SE domain. Therefore, the Synonyms Replacement component leverages an SE thesaurus, which is a word2Vec model trained on Stack Overflow posts to capture the SE terms~\cite{Efstathiou_2018MSR}. The SE thesaurus returns `issue' and `error' as synonyms to the term `bug'.

This component creates a new candidate query by replacing one token within a query from the original training set with its synonym. In case a query has two or more tokens to be replaced, the Synonyms Replacement component generates  new candidate queries based on all the possible combinations of the replaceable tokens' synonyms. In the working example in Figure~\ref{fig:working_example}, the Synonyms Replacement component replaces the verb token `cause' from the query ``What files cause the most issues?'' with its synonyms `induce' and `generate', thus creating two new candidate queries (e.g., ``What files induce the most issues?''). 

In our preliminary analysis, we find that replacing nouns with their synonyms generates too much noise. For example, the synonym of the `developer' and `button' tokens are replaced with `prismic' and `buttom'; respectively. Thus, we opt to just replace verb tokens with their synonyms in our study.

\underline{\textit{Paraphrasing:}} One aspect of expanding the training dataset is to expose the NLU to new terminologies. The other important aspect is training NLU on a variety of sentence structures for queries. Chatbot developers typically paraphrase the queries they add to the training set because users can phrase the same question in different ways~\cite{Abdellatif_MSR2020}. In fact, this process is recommended by the NLU vendors to enhance their performance in intent classification~\cite{sentenceStructureLUIS_link,sentenceStructureRasa_link}.

Therefore, the Paraphrasing component diversifies the sentence structure of candidate queries while preserving their meaning (intent). 

The paraphrasing component takes as input each candidate queries from the Synonyms Replacement component. 
To paraphrase queries, this component leverages the recent transformer based neural machine translation (Seq2Seq) model called BART~\cite{lewis_2020ACL}. BART is a general language model proposed by Facebook AI and has been used by prior work for paraphrasing task~\cite{Zhou_2020ACL,Dopierre_2021ACL,West_2021ACL} as it achieves state-of-art performance in various NLP tasks (e.g., machine translation, summarization, and text generation)~\cite{lewis_2020ACL}. BART is trained through corrupting the input example (e.g., delete one of its tokens) during the training stage and then predicting the correct form of the corrupted sentence. We fine-tune BART to perform paraphrasing task (discussed in Section~\ref{sec:casesetup6}).

In the working example (Figure~\ref{fig:working_example}), the Paraphrasing component takes the two candidate queries from the Synonyms Replacement component as an input and outputs paraphrased queries (e.g., ``Most issue inducing files?''). 
The final output of the Paraphrasing component is a list of new candidate queries that preserve the intent and have both new terms and different sentence structures compared to the original training set.

\underline{\textit{Diversity Filter:}} The main goal of the augmentation approach is to generate candidate queries with the highest potential of improving the NLU's performance. Therefore, the Diversity Filter selects the candidate queries yielded by the Paraphrasing component that are most syntactically different compared to the original training set. This helps to mitigate the issue of overfitting the NLU that may occur when the candidate queries do not increase the syntactical diversity of the original training set. 

As a means to measure the diversity, this component computes the Levenshtein distance~\cite{levenshtein1966SPD} (number of edits between two queries) between the candidate queries and the original training set. The higher the Levenshtein distance, the more dissimilar the queries. In the working example (Figure~\ref{fig:working_example}), the Levenshtein distance between the candidate query ``Most issue inducing files?'' and the original training query ``What files contain the most issues?'' is 14. 

Then, the Diversity Filter ranks the candidate queries based on their minimum Levenshtein distance to any query in the original training set. In other words, the Diversity Filter ranks candidate queries by the highest minimum Levenshtein distance from the original query, placing queries that are more syntactically different from the original query at the top. This approach balances the inclusion of queries that are syntactically diverse but still convey the same semantic intent as the original query.
Finally, the top $N$ candidate queries are kept by the Diversity Filter while the rest are discarded ($N$ is configurable). In other words, the Diversity Filter component discards candidate queries that are syntactically similar to those that are already present in the original training set. For instance, if we set \textit{N}=1 in our showcase example, the candidate query ``Most issue inducing files?'' passes the filter because it has the highest minimum Levenshtein distance (12).

\underline{\textit{Entity Labeler:}} Typically, the chatbot training datasets include annotations of both intents and entities for all queries in the set. Such annotations are essential for some NLUs for the intents classification step~\cite{Abdellatif_2021TSE}. However, the candidate queries that are retained after the Diversity Filter do not contain any entity annotations. Hence, the Entity Labeler component uses heuristics to label the entities in the candidate queries. To establish the heuristics, we examined 400 random samples from different intents generated by the Paraphrasing component and found that the entities remain the same or experience minor modifications only during the paraphrasing. For example, the FileName entity (e.g., `ConsumerRecords') could be changed (e.g., `Consumer Records') during the paraphrasing. The only exception here is the DateTime entities, where a specific date (e.g., 21-12-2022) can be changed to `last week'. Based on our observations, we define heuristics to label entities in the candidate queries. Therefore, the Entity Labeler component reads all labeled entities in the original training dataset and automatically labels the entities in the candidate queries using the defined heuristics.

\underline{\textit{Data Merger:}} The output of any augmentation approach should be a training dataset that is ready for use. The Data Merger component is responsible for adding the output queries of the augmentation approach to their corresponding intents in the original training set to generate augmented training set file. Therefore, chatbot practitioners can use the output file to train the NLU used in their chatbots.

\section{Evaluation Setup}
\label{sec:casesetup6}

In this section, we present the setup of our case study to evaluate the impact of the proposed augmentation approach on the NLU's performance. We detail our selection of the SE datasets used in the evaluation, NLU platform used for training and testing, tuning the BART transformer, and experiment design.

\subsection{Datasets}
\label{subsec:dataset}

\begin{table*}[t!]
\centering
\caption{Performance comparison results for augmentation approach against the baseline and human.}
\vspace{-0.15in}
\label{table:dataset}
\hspace*{-1cm}
\begin{tabular}{llp{8cm}lll}
\hline
\textbf{Dataset} & \textbf{Intent} & \textbf{Definition} & \textbf{Train} & \textbf{Test} & \textbf{Total} \\ \hline
\parbox[t]{2mm}{\multirow{10}{*}{\rotatebox[origin=c]{90}{\textbf{Repository}}}} & \textbf{BuggyCommitsByDate} & Present the buggy commit(s) which happened during a specific time period. & 66 & 13 & 79 \\
 & \textbf{BuggyCommit} & Identify the bugs that are introduced because of certain commits. & 52 & 9 & 61 \\
 & \textbf{BuggyFiles} & Determine the most buggy files in the repository to refactor them. & 37 & 13 & 50 \\
 & \textbf{FixCommit} & Identify the commit(s) which fix a specific bug. & 31 & 11 & 42 \\
 & \textbf{BuggyFixCommits} & Identify the fixing commits that introduce bugs at a particular time & 32 & 7 & 39 \\
 & \textbf{CountCommitsByDates} & Identify the number of commits that were pushed during a specific time period. & 11 & 21 & 32 \\
 & \textbf{ExperiencedDevFixBugs} & Identify the developer(s) who have experience in fixing bugs related to a specific file. & 15 & 14 & 29 \\
 & \textbf{OverloadedDev} & Determine the overloaded developer(s) with the highest number of unresolved bugs. & 15 & 9 & 24 \\
 & \textbf{FileCommits} & View details about the changes that occurred on a file. & 10 & 12 & 22 \\
 & \textbf{CommitsByDate} & Present the commit information (e.g., commit message) at a specific time. & 8 & 12 & 20 \\ \hline
\parbox[t]{2mm}{\multirow{5}{*}{\rotatebox[origin=c]{90}{\textbf{Ask Ubuntu   }}}}& \textbf{SoftwareRecommendation} & Looking for applications that perform specific task (e.g., extract images from PDF). & 17 & 40 & 57 \\
 & \textbf{MakeUpdate} & Looking for information related to upgrading Ubuntu version to a newer version. & 10 & 37 & 47 \\
 & \textbf{ShutdownComputer} & Related fix shutdown issues in Ubuntu OS. & 13 & 14 & 27 \\
 & \textbf{SetupPrinter} & Setup a printer and fix printer installation issues. & 10 & 13 & 23 \\ \hline
\parbox[t]{2mm}{\multirow{6}{*}{\rotatebox[origin=c]{90}{\textbf{Stack Overflow     }}}} & \textbf{LookingForCodeSample} & Looking for information related to implementation (e.g., code snippets). & 66 & 66 & 132 \\
 & \textbf{UsingMethodImproperly} & An improper use of a method is causing unexpected behaviour. & 25 & 26 & 51 \\
 & \textbf{LookingForBestPractice} & Looking for the recommended (best) practice, approach or solution for a problem. & 6 & 6 & 12 \\
 & \textbf{FacingError} & Facing an error or a failure in a program, mostly in the form of an error message. & 5 & 5 & 10 \\
 & \textbf{PassingData} & Passing data between different frameworks or method calls. & 5 & 5 & 10 \\ \hline
\end{tabular}
\vspace{-0.15in}
\end{table*}

To evaluate how effective the approach is in augmenting a variety of SE datasets, we select three distinct  datasets: Repository~\cite{Abdellatif2020EMSE}, Ask Ubuntu~\cite{braun_2017ACL}, and Stack Overflow~\cite{Ye_2016SANER}. Our dataset selection is based on three criteria. First, these datasets represent realistic questions that software practitioners ask about software projects and development. Second, they have adequate numbers of training and testing queries (ten or more queries per intent) to conduct proper evaluation. Table 1 presents the intents in each of the datasets, their definitions, and the distribution of queries in the training and test sets corresponding to each intent. Finally, the datasets are publicly available and have been used in previous studies~\cite{Abdellatif_2021TSE,Larson_2019ACL,shridhar_2020ACL}. In the following, we provide a description of each dataset.
\\

\noindent\textbf{Repository:} Contains questions asked by software practitioners about their software repositories to the MSRBot~\cite{Abdellatif2020EMSE}. One example query from this dataset is ``What are the commits that introduce bug HHH8492?''. For this dataset, the MSRBot developers created the training set manually and composed the test set from queries asked by the MSRBot users. Thus, the training and test sets in this corpus originate from a real-life use case of an SE-based chatbot in practice. The Repository dataset contains 398 queries belonging to ten intents in total.

\noindent\textbf{Ask Ubuntu:} This dataset was constructed using the most popular posts from the Ubuntu Q\&A community on Stack Exchange, one of the most popular online discussion forums~\cite{braun_2017ACL}. \citet{braun_2017ACL} selected the most popular questions, which were then annotated using Amazon Mechanical Turk. An example of a query from this dataset is ``How to upgrade Ubuntu 14.04.1 to 14.04.2''. This dataset contains 154 queries split into four intents. It is important to note that we discarded the `Other' intent because it had an insufficient number of queries (i.e., three queries) for our evaluation.

\noindent\textbf{Stack Overflow:} Contains labeled software development questions from Stack Overflow~\cite{Ye_2016SANER}, another popular development Q\&A website. The queries in this dataset were collected by \citet{Ye_2016SANER} and then \citet{Abdellatif_2021TSE} labeled the queries’ intents. In total, this dataset is composed of 215 queries and five different intents. One example query from this dataset is ``How can I get Font X offset width in java2D?''.

\subsection{NLU} 
The goal of the augmentation approach is to improve the NLU's performance by augmenting a given training dataset. Hence, we need to select an NLU platform to train its model and perform our evaluation. For this study, we select Rasa, an open-source NLU platform developed by Rasa Technologies~\cite{Conversa24:online}. Unlike third-party NLUs that operate on the cloud (e.g., Google Dialogflow), Rasa can be installed, configured, and run locally, which consumes fewer resources. And the Rasa implementation stays the same during our experiment, while the internal implementation of third-party NLUs might change without any prior notice to the users~\cite{Abdellatif_2021TSE}. Moreover, Rasa has been used by prior work to develop SE chatbots~\cite{Lin_2020BotSE,Abdellatif2020EMSE,Dominic_2020BotSE}. In our implementation, we used Rasa version 2.5 as it was the latest stable version available at the time of our study.

\begin{table*}[tbh!]

\caption{Performance comparison results for augmentation approach against the baseline and human.}
\vspace{-0.15in}
\small

\label{tab:RQ2-resutls}
\centering
\begin{tabular}{@{}c|c|r|rrr|rr@{}}
\toprule
\multirow{2}{*}{\textbf{Dataset}}        & \multirow{2}{*}{\textbf{\begin{tabular}[c]{@{}c@{}}No. of \\ Queries/Intent\end{tabular}}} & \multicolumn{1}{l|}{\textbf{Baseline}} & \multicolumn{3}{c|}{\textbf{Augmentation Approach}}            & \multicolumn{2}{c}{\textbf{Human}}   \\
 &   & \multicolumn{1}{l|}{\textbf{F1-score}} & \multicolumn{1}{l}{\textbf{F1-score}} & \multicolumn{1}{l}{\textbf{\% Improvement}} & \multicolumn{1}{l|}{\textbf{\% Optimal}} & \multicolumn{1}{l}{\textbf{F1-score}} & \multicolumn{1}{l}{\textbf{\% Improvement}} \\
\midrule
\multirow{3}{*}{\textbf{Repository}}     & \textbf{One} & 43.7 & 44.7* & 2.3 & 6.9 & 58.1* & 33.0  \\
         & \textbf{Three}                     & 62.8 & 64.3* & 2.4 & 34.1 & 67.2* & 7.0 \\
         & \textbf{Five}                     & 66.4 & 68.5* & 3.2 & 60.0 & 69.9* & 5.3  \\
         \midrule
\multirow{3}{*}{\begin{tabular}[c]{@{}c@{}}\textbf{Ask} \\ \textbf{Ubuntu}\end{tabular}}     & \textbf{One} & 71.6 & 73.5* & 2.7 & 17.3 & 82.6* & 15.4  \\
         & \textbf{Three}                     & 84.1 & 83.6* & -0.6 & - & 90.4* & 7.5  \\
         & \textbf{Five}                     & 87.1 & 84.2* & -3.3 & - & 90.2* & 3.6  \\
         \midrule
\multirow{2}{*}{\begin{tabular}[c]{@{}c@{}}\textbf{Stack} \\ \textbf{Overflow}\end{tabular}} & \textbf{One} & 32 & 32.2 & 0.6 & 2.5 & 40* & 25  \\
         & \textbf{Three}                     & 36.9 & 37.9* & 2.7 & 12.3 & 45* & 22.0   \\
         \bottomrule
\end{tabular}
	\begin{tablenotes}[flushcenter]
 \footnotesize 
 \addtolength{\itemindent}{1.2cm}
		\item * The difference is statistically significant (p-value<0.05) compared to Baseline.
	\end{tablenotes}
 \vspace{-0.1in}
\end{table*}

\subsection{BART Tuning}
As discussed in Section~\ref{sec:approach6}, the paraphrasing component uses BART transformer to paraphrase the candidate queries resulted from the Synonyms Replacement component. BART is trained on 160GB of documents (e.g. Wikipedia, news articles, stories) with a sentence reconstruction loss~\cite{lewis_2020ACL}. To use BART, we need to fine-tune it to the specific task at hand which is, in our case, to paraphrase text. Therefore, we use the following three datasets to fine-tune BART: 1) Quora Question Pairs contains over 149,263 lines of potential duplicate pairs of questions obtained from Quora social Q\&A website, 2) Microsoft Research Paraphrase composed of 3,749 pairs of sentences extracted from the internet (e.g., news sources) and then annotated by humans to indicate whether each pair captures the same semantics, and 3) Paraphrase Adversaries from Word Scrambling contains 25,368 pairs of paraphrased sentences generated using word swapping and back-translation created by \citet{Zhang_2019naacl}. These three datasets have been used in prior work for text paraphrasing~\cite{Dopierre_2021ACL,West_2021ACL}, which makes them a solid choice to fine-tune BART. Furthermore, we use the \textit{BART-large} model, which has 12 layers in the encoder and decoder, and more than 374 million of parameters. We train the BART model on a cloud with 6 core Intel E5-2683 v4 Broadwell, 64GB of RAM, and NVIDIA V100 Volta GPU (32G HBM2 memory). We examined BART's output with different numbers of returned paraphrases (i.e., 3, 5, 7, and 10) and found that BART performs best in terms of having diverse sentence structure and preserving the semantics of the input when it returns three paraphrases at most for any input query.

\subsection{Evaluation Settings}
\label{sec:evaluation-settings}
To evaluate the impact of using the augmentation approach on the NLU's performance, we train the NLU after augmenting the original training set and then evaluate the NLU's performance using the test set.
Unlike the Repository and Ask Ubuntu datasets, there is no predefined train-test split in the Stack Overflow dataset. Therefore, we divide the Stack Overflow dataset into 50\%-50\% for training and test splits through random stratified sampling. 
Table~\ref{table:dataset} presents the distribution of queries for each intent in the training and test splits in the Repository, Ask Ubuntu, and Stack Overflow datasets. It is important to note that we repeat this step 10 times to obtain different inputs (training queries) and evaluate the performance of the augmentation approach on diverse input queries.

After obtaining the training and test splits for all datasets, we craft three scenarios: A scenario with 1 query/intent, 3 queries/intent, and 5 queries/intent. The goal of our scenarios is to evaluate the potential of the augmentation approach in scenarios with little training dataset where augmentation have the tendency to yield the biggest impact on the NLU's performance.
For each scenario, we randomly select the queries per intent from the original training dataset.
For example, the Stack Overflow dataset has five intents. Thus, the 3 queries/intent scenario in the Stack Overflow dataset has in total of 15 training queries (i.e., 3 queries per intent).

\subsection{Performance Evaluation}
To assess the NLU's performance in classifying intents, we compute the widely used metrics of precision, recall, and F1-score. Precision is the percentage of the correctly classified queries to the total number of classified queries for that intent (i.e., Precision = $\frac{TP}{TP+FP}$). The recall is calculated as the percentage of the correctly classified queries to the total number of queries for that intent in the test set (i.e., Recall = $\frac{TP}{TP+FN}$). To have an overall score, we combine both the precision and recall using the F1-score weighted by class' support (weighted F1-score), which has been used in similar studies~\cite{Abdellatif_2021TSE,Barash_2019FSE,Ilmania_2018IALP}. More specifically, we start by computing the F1-score (i.e., F1-score = \(2 \times \frac{Precision \times Recall}{Precision + Recall}\)) for all classes. Then, we aggregate all F1-scores using the weighted average, with the class' support as weights. It is important to note that, although we evaluate all three metrics, we only present the weighted F1-score in the paper. We make the precision, recall, and F1-score results for each intent publicly available online~\cite{Augmenti93:online}.

\section{Results}
\label{sec:results6}
In this section, we present the results of our case study with respect to the three research questions. For each research question, we present the motivation for the question, detail the approach to answer the question, and present the results.

\subsection{RQ1: Can the augmentation approach improve the NLU's performance?}

\textbf{Motivation:} 
Previous studies show that augmenting the training set used to train the NLU leads to improving its performance~\cite{Abdellatif_2021TSE}. However, the process of crafting and collecting more training data is done manually, which is a costly and time-consuming task~\cite{Dominic_2020BotSE,Abdellatif2020EMSE}. The goal of this research question is to evaluate the impact of using the augmentation approach (discussed in Section~\ref{sec:approach6}) on the NLU's performance. Succeeding in this task saves time and resources, which allows practitioners to focus on the critical tasks of their chatbots rather than augmenting the training set.

\begin{table*}[]
\small
\centering
\caption{Sample of the augmented queries by the augmentation approach.}
\vspace{-0.15in}
\label{tab:AugmentedExamplesSample}
\scalebox{0.95}{
\begin{tabularx}{\linewidth}{XXl}  
\toprule
\textbf{Original Query} & \textbf{Augmented Queries} & \textbf{Intent} \\
\midrule
Show me the number of commits happened last week & How many commits did you commit between last week and this week? & CommitsCountByDate \\
\midrule
Show me the classes which introduced the most of bug &  What files are the ones that added most of bugs & BuggyFiles \\
 \midrule
How can one shutdown a PC using the keyboard? & Hotkey to shut down from login screen? & ShutdownComputer \\
\midrule
Python inserting variable string as file name & How do I insert a string as a file name in Python? & LookingForCodeSample \\
\midrule
Spring 4.0.2 schema error & What is the reason behind the schema error in Spring 4.0.2? & FacingError \\
\bottomrule
\end{tabularx}}
 \vspace{-0.1in}
\end{table*}

\noindent\textbf{Approach:} 
To answer this RQ and put the results of the augmentation approach into perspective, we perform three different experiments:

\noindent\textbf{(1) Baseline}: Establishes a baseline for the NLU's performance. In this experiment, we use the original queries from each of the scenarios (i.e., 1, 3, and 5 queries/intent) to train the NLU. Hence, no augmented query is included in the NLU training. 

\noindent\textbf{(2) Augmentation Approach}: This experiment reflects the situation where a chatbot practitioner augments the original training dataset using our augmentation approach. In this experiment, we apply the augmentation approach to scenarios 1, 3, and 5 queries/intent. Then, we augment the queries resulting from the augmentation approach into the scenario set. In our study, we set the configuration to augment one query per intent \textit{N}=1 such that the Diversity Filter keeps only the top-ranking candidate query. For example, augmenting a scenario of 1 query/intent using the augmentation approach yields a training set of 2 queries/intent. This makes our evaluation manageable since it requires manually examining all the generated queries for all intents across the three different datasets to obtain insights about the augmented queries. Also, it makes our study manageable in terms of consumption power since our approach requires running two deep learning models, BART transformer and NLU, to generate queries.

\noindent\textbf{(3) Human}: In this experiment, we evaluate the impact of using queries crafted by humans to augment the original training set on the NLU's performance. This experiment reflects a situation where a chatbot practitioner starts with a set of training queries and then manually augments the training set. The human-augmented queries are high-quality and can most likely improve the NLU's performance compared to any augmentation approach. In this experiment, from the dataset discussed in Section 4.1, we randomly select one query per intent from the training split that was not used in the scenarios (e.g., 3 queries per intent) and augment the selected query to the scenario. Thus, we have the same settings as the Augmentation Approach experiment (i.e., augmenting one query per intent).

We follow the same evaluation steps for all datasets, experiments, and scenarios. First, we train the NLU using the scenario's queries resulting from the experiment. Next, we use the test set to evaluate the NLU's performance for each experiment and record the results. It is important to note that in the Stack Overflow dataset, we only run our evaluation on scenarios 1 query/intent and 3 queries/intent. This is because some intents (e.g., `PassingData') in the 5 queries/intent scenario of the Stack Overflow dataset do not have enough training queries left to be used as input queries to randomly select in the Human experiment.

To measure the improvement in the NLU's performance achieved by a specific experiment (i.e., Augmentation Approach and Human) to the baseline, we resort to the \%Improvement metric. The \%Improvement for an experiment (EXP) is measured using the equation:\\

\small $\% Improvement_{(Exp)} = $ $\frac{F1\,Score_{(Exp)} - F1\,Score_{(Baseline)}}{F1\,Score_{(Baseline)}} * 100$.\normalsize\\

Since the main idea of the augmentation approach is to imitate the chatbot practitioner on how they augment their training datasets, we use \%Optimal metric to measure how close the augmentation approach is to the human augmentation performance. This measure calculates the ratio of the \%Improvement achieved by the Augmentation approach to the \%Improvement achieved by the Human experiment as follows:\\
\begin{center}
$\% Optimal = $ $\frac{\% Improvement_{(Augmentation\,Approach)}}{\% Improvement_{(Human)}} * 100$\\
\end{center}
It is worth noting that we repeat the evaluation step 100 times for each scenario and experiment to reduce the randomness effect of the NLU's model. We report the average results from those runs in our results. We use the non-parametric Mann-Whitney U test~\cite{wilks2011statistical} to statistically compare the difference between the results of all experiments. We chose the Mann-Whitney U test because it does not assume any specific distribution of the data.

\begin{figure*}[t!]
     \centering
     \begin{subfigure}[b]{0.33\textwidth}
         \centering
         \includegraphics[width=\textwidth]{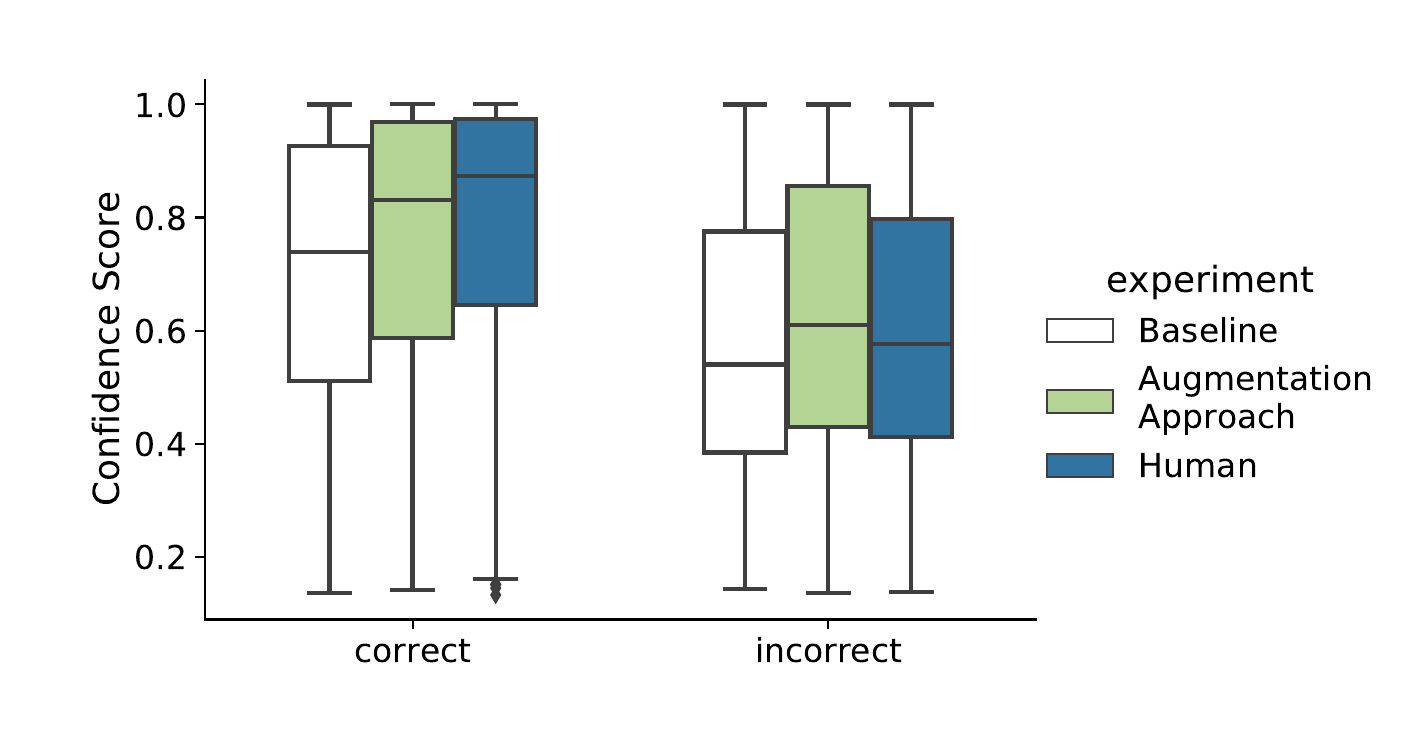}
         \caption{1 query/intent}
         \label{fig:repo_cs_1}
     \end{subfigure}
     \hfill
     \begin{subfigure}[b]{0.33\textwidth}
         \centering
         \includegraphics[width=\textwidth]{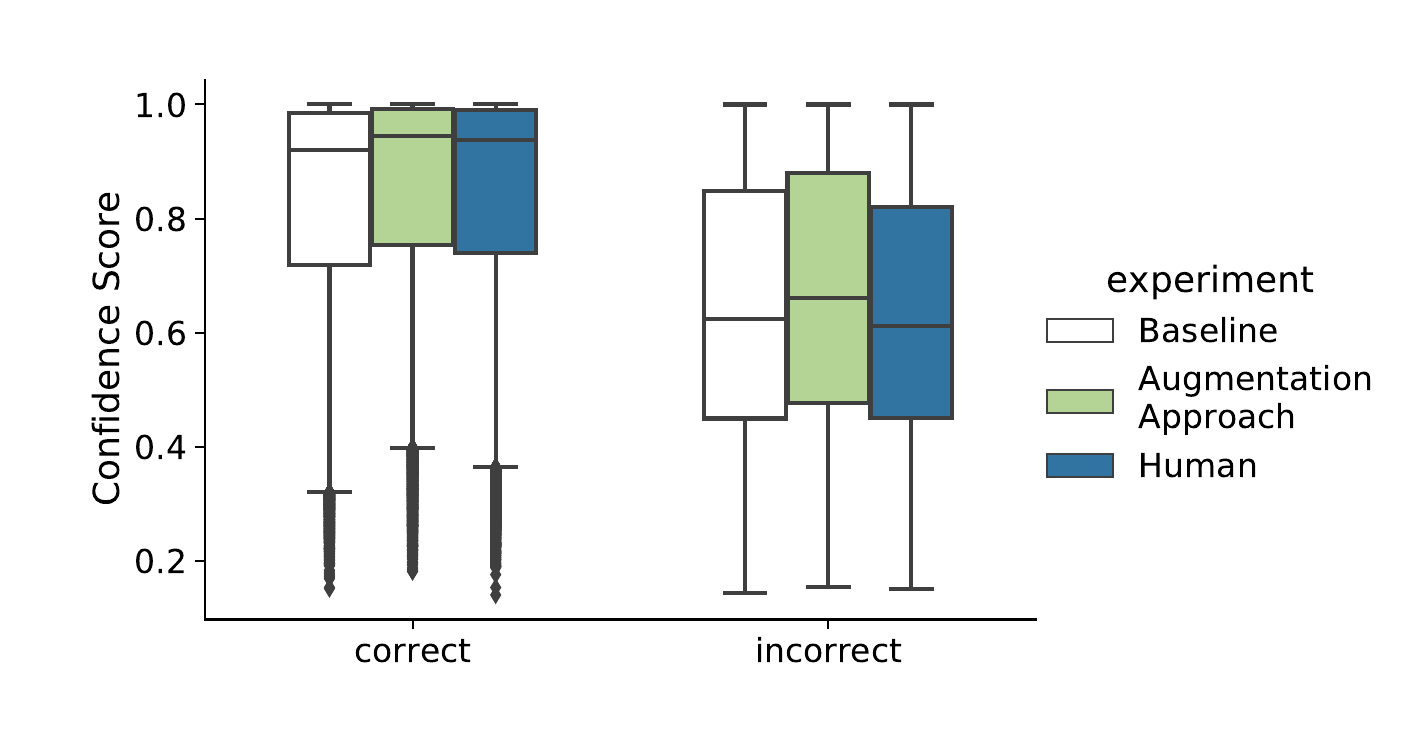}
         \caption{3 queries/intent}
         \label{fig:repo_cs_3}
     \end{subfigure}
     \hfill
     \begin{subfigure}[b]{0.31\textwidth}
         \centering
         \includegraphics[width=\textwidth]{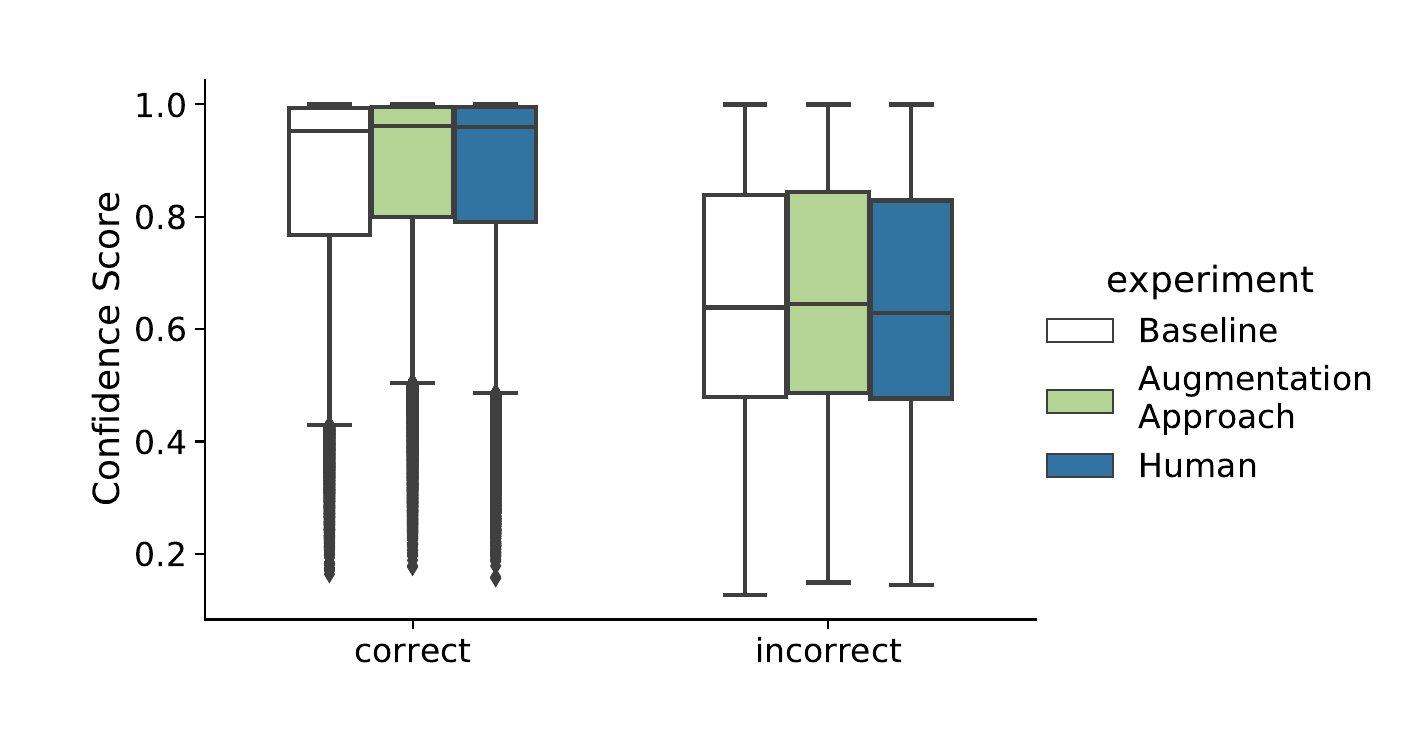}
         \caption{5 queries/intent}
         \label{fig:repo_cs_5}
     \end{subfigure}
     \vspace{-0.15in}
        \caption{The confidence score distributions for scenarios 1, 3, and 5 queries/intent in the Repository dataset.}
        \label{fig:RQ3_results}
        
\end{figure*}

\noindent\textbf{Results:}
Table~\ref{tab:RQ2-resutls} presents the F1-scores, \%Improvement, and \%Optimal for Baseline, Augmentation Approach, and Human experiments on all scenarios and datasets. From the table, we observe that the augmentation approach marginally improves the NLU's performance (\%Improvement $>$ 2\%) in six scenarios. On the other hand, there are two scenarios where the augmentation approach does not improve the performance of the NLU (\%Improvement $\leq$ 0). For example, applying the augmentation approach on the scenario of 1 query/intent in the Ask Ubuntu dataset decreases the NLU's performance (\%Improvement $<$ -0.06). Among the six scenarios where there is an improvement in the NLU's performance, we notice that the augmentation approach is most effective in the scenarios that have more queries. For example, the augmentation approach achieves similar results to Human in 5 queries/intent scenario for the Repository dataset with \%Optimal of 60\%. This is because those scenarios (e.g., 5 queries/intent) have more queries that are used to augment more diverse queries compared to the scenarios with fewer queries (e.g., 1 query/intent).

Upon closer examination of the augmented queries across all scenarios and datasets, we have two main observations: 1) Some of the augmented queries have different sentence structures than the queries in the original training set. Table~\ref{tab:AugmentedExamplesSample} presents a sample of the original training queries and their corresponding queries augmented through the augmentation approach. For instance, the augmentation approach augments the query ``Show me the number of commits happened last week'' from the original training set to a new candidate query ``How many commits did you commit between last week and this week?''. In some cases, the augmented queries make minor changes to the original queries. For example, the augmentation approach replaces the word `fix' with `remedy' in the initial training query ``Who has the most bugs to fix?'' to generate a new candidate query ``Who has the most bugs to remedy?''. 2) Upon manual examination of all augmented queries across different scenarios and datasets, we find that the augmentation approach preserves the semantic (intent) of the original queries as shown in Table~\ref{tab:AugmentedExamplesSample}. We further examined all generated queries (4,380 queries) by Paraphrasing component to investigate whether BART transformer preserves the intent of the original queries before passing the queries to the Diversity Filter component. We find that 99.4\% of paraphrased queries maintain the intent of their original queries, with only 0.6\% (26 out of 4,380) of queries having a changed intent across all runs in different scenarios and datasets.
For example, the initial training query ``Give me the changes of ConnectRestConfigurable file'' asks about presenting the changes that touch ConnectRestConfigurable file (FileCommits intent), however, the augmented query is ``How can I change the changes of ConnectRestConfigurable file'', which changes the meaning of the query to ask about modifying the commits that changed the file.

To better understand the factors that the augmentation approach yields to low performance for scenarios with 3 and 5 queries/intent in the Ask Ubuntu dataset, we examined the augmented queries of all runs in those scenarios. Surprisingly, we find that the augmentation approach often augments queries with small modifications to the initial training query. For example, the approach modified the Ubuntu version in ``How to upgrade Ubuntu [14.04.1](UbuntuVersion) to [14.04.2](UbuntuVersion)?'' to augment a new query ``How to upgrade from Ubuntu [10.10](UbuntuVersion) to [11.04](UbuntuVersion)?''. In other cases, the approach omits some text from the initial training queries. For example, in the ``How To Install [Canon LBP2900B](Printer) printer in [14.04 LTS](UbuntuVersion)? I tried the method for LBP2900 but it didnt work'' query, the approach removed ``I tried the method for LBP2900 but it didnt work'' to create the augmented query (i.e., ``How To Install [Canon LBP2900B](Printer) printer in [14.04 LTS](UbuntuVersion)? I tried the method for LBP2900 but it didnt work''). Thus, the high similarity between the augmented queries and the original training queries might cause overfitting and decrease the performance of the NLU. On the other hand, we observe that the performance for 3 and 5 queries/intent in the Ask Ubuntu dataset is high (F1-score > 84\%), indicating that the augmentation approach is particularly beneficial in situations where the performance of NLU is low, which is the main goal of our study (i.e., improve the low performance of the NLU).

Our study results show that the augmentation approach has a small, yet positive impact on the performance of the NLU.
We reiterate that this improvement is achieved through automation and does not require human input. Additionally, we believe that the generated queries by the approach could serve as inspiration for chatbot practitioners to diversify sentence structures and synonyms while manually expanding the training dataset.

\conclusion{Augmented queries provide small but significant improvement in the NLU's performance.
In various scenarios, the approach is most efficient when augmenting training datasets of 3 and 5 queries/intent. Additionally, it augments queries with varying sentence structures while preserving their original semantics.}

\subsection{RQ2: Does the augmentation approach increase the NLU's confidence in its classification?}
\label{sec:RQ3}

\textbf{Motivation:} 
Each intent classification performed by the NLU has a corresponding confidence score, as discussed in Section~\ref{sec:background6}. Chatbot developers rely on the confidence score returned with the classified intent to determine the chatbot's next action~\cite{Abdellatif_2021TSE}. More specifically, developers tend to design their chatbots so that, if the NLU is not confident in its intent classification, the chatbot asks for further clarification from the user (i.e., ``Sorry, I did not understand your question, could you please rephrase it?''). Developers expect a well-trained NLU to provide higher confidence scores for correctly classified intents and lower scores for misclassified intents to minimize wrong actions performed by the chatbot. One way to increase the NLU's confidence in its classification is to train the NLU on more queries for each intent (i.e., augment the training set). Therefore, in this RQ, we evaluate the impact of using the augmentation approach on the NLU's confidence, particularly with regard to the confidence scores returned with the correctly and incorrectly classified intents.

\noindent\textbf{Approach:} 
To answer this RQ, we use the output of the three experiments (Baseline, Augmentation Approach, and Human) described in RQ1. In particular, we examine the confidence scores returned by the NLU for the test set queries across all the experiments. By using the three experiments, we establish a baseline (Baseline experiment) for how confident the NLU is in its intent classification. Then, we use that baseline to measure the impact of using the augmentation approach on the NLU's confidence (Augmentation experiment) compared to the impact of using human-augmented queries (Human experiment). We hypothesize that augmenting more queries increases the NLU's confidence scores for the correctly classified intents while decreasing its confidence for the misclassified intents. This is because the NLU will be exposed to more ways (i.e., different syntactic structures) of asking about the same intent. Finally, to compare the confidence scores for the correctly and incorrectly classified intents, we present the distributions of confidence scores for each case. Also, we perform the non-parametric unpaired Mann-Whitney U test~\cite{wilks2011statistical} to verify whether the difference in the confidence scores across the experiments' results (e.g., Baseline vs Augmentation Approach experiments) are statistically significant.

\noindent\textbf{Results:} 
Figure~\ref{fig:RQ3_results} shows the confidence score distributions for both the correctly and incorrectly classified intents for all scenarios in the Repository dataset. As shown in the figure, in all experiments, we find that the median confidence scores of correctly classified intents are higher than the misclassified intents. In fact, these results are in-line with the ones in the prior work~\cite{Abdellatif_2021TSE}.

From the figure, we also observe that using the Augmentation Approach significantly increases the NLU's confidence in its intent classification for correctly classified intents, particularly in the 1 query/intent scenario, compared to the Baseline across all scenarios. 
For example, in 1 query/intent scenario, the Augmentation approach increases the median confidence scores (76.3\%) by 6\% compared to the Baseline (70.1\%). In some cases, the Augmentation Approach outperforms the Human experiment in terms of confidence scores for correctly classified intents. For example, in the 3 queries/intent scenario, the Baseline experiment has a median confidence score of 83.1\%, while the Augmentation approach and Human experiment have median confidence scores of 85\% and 84.3\%, respectively. 

For incorrectly classified intents, we find that the augmentation approach increases the NLU's confidence on misclassified queries compared to the Baseline. That is, the median confidence scores for incorrectly classified intents is higher when training the NLU with the augmented dataset. 
Surprisingly, using human-augmented queries (i.e., Human experiment) increases the overall confidence scores for misclassified intents compared to the baseline. For example, in the 1 query/intent scenario, the median confidence score is 58.1\% for the Baseline, while the median confidence scores for the Augmentation approach and Human experiments are 63.3\% and 60.3\%, respectively.

\conclusion{Using the Augmentation Approach increases the NLU's confidence in intent classification for correctly classified intents across scenarios. Nonetheless, the augmented queries in the Human and Augmentation Approach experiments increase the NLU's confidence in misclassified intents.}

\section{Discussion}
\label{sec:lessonslearned6}

\textbf{Quality of generated data is far more important than quantity.}
NLU achieves reasonable performance even when trained on a few queries per intent, as discussed in RQ1. Results show that with only one training query per intent (i.e., 1 query/intent scenario), the NLU reaches an F1-Score ranging from 32\% in the Stack Overflow and up to 71.6\% in the Ask Ubuntu datasets (see Table~\ref{tab:RQ2-resutls}) in the Baseline experiment. This observation is further reinforced when looking at the baseline performance with 5 queries/intent scenario, where the NLU achieves an F1-score up to 87\% in the Ask Ubuntu dataset. Although this seems promising for chatbot developers, achieving a robust NLU performance for real-scenario is still very challenging, as chatbots may deal with sensitive data (e.g., software project data) or performs critical tasks (e.g., merging vulnerable code into the main branch). Given this level of robustness of modern NLUs, augmentation approaches should focus on the quality of the generated queries more than just its ability to increase the size of the training dataset. Furthermore, having high-quality augmented queries inspires chatbot practitioners to add a variety of new queries to their dataset, ultimately resulting in improved performance.

\textbf{Practitioners need to consider fine-tuning the confidence threshold of their chatbots.}
The confidence score determines the chatbot's response, with high scores resulting in action (e.g., answering the user's question) and low scores prompting the chatbot to request clarification from the user in order to improve its understanding of the user's intent~\cite{Abdellatif_2021TSE}. The Augmentation approaches raise the confidence levels of NLUs, both for correctly and incorrectly classified intents, as discussed in RQ2.
Therefore, we recommend practitioners who use augmentation approaches consider fine-tuning the appropriate confidence score threshold of their chatbots to accept the classified intent or trigger the fallback action. Moreover, our results suggest that future augmentation approaches should take into account the confidence scores for both correctly and incorrectly classified intents when augmenting queries.

\textbf{There is a need for a well-crafted dataset of paraphrased queries for the SE domain.}  
In our study, we configured BART to return three paraphrases, as discussed in Section~\ref{sec:casesetup6}. In RQ1 results, we observe that BART tends to generate queries with only slight variations between them. For example, the ``How to move content from QListWidget to a QStringList with PyQt4?'' query in the Stack Overflow dataset, BART generates ``How to moving content from QListWidget to a QStringList with PyQt4?'', ``How to moves content from QListWidget to a QStringList with PyQt4?'', and ``How do I move content from QListWidget to a QStringList with PyQt4?'', which have high similarities between the queries. We reiterate that in fine-tuning the BART model to paraphrase queries, we reuse three datasets (e.g., Quora question pairs) that have been used in prior work~\cite{Dopierre_2021ACL,West_2021ACL}. None of these datasets are related to the SE domain. The lack of a more specialized SE dataset might limit BART's ability to generate paraphrased queries specific to SE. To the best of our knowledge, there is no crafted dataset that contains pairs of paraphrased queries related to the SE domain. We plan (and encourage others) to create benchmarks that contain paraphrased pairs of queries representing different SE tasks in order to fully utilize the potential transformers in the SE context.

that only one intent had new examples added to it. However, we also observed that the performance of other intents dropped at the same time. And while this can potentially be due to the randomness of the model, it can also indicate that adding examples to one intent can affect the performance of other intents.

\section{Related Work}
\label{sec:relatedwork}
The goal of this paper is to propose an approach to augment the training datasets for the SE chatbots. Thus, we divide the prior work into two main areas; work related to developing SE chatbots and work related to dataset augmentation.

\noindent\textbf{Software Engineering Chatbots.}
A plethora of studies propose chatbots to assist developers in their daily tasks~\cite{Dominic_2020BotSE,Bradley_2018ICSE,Paikari_2019BotSE,Abdellatif2020EMSE}. For example, Bradley et al.~\cite{Bradley_2018ICSE} propose Devy, a chatbot to help software developers in their development tasks (e.g., pushing the new code changes to the project repository). Dominic et al.~\cite{Dominic_2020BotSE} develop a chatbot to help newcomers with the onboarding process on OSS projects. Abdellatif et al.~\cite{Abdellatif2020EMSE} develop a chatbot to answer questions related to software projects. Paikari et al.~\cite{Paikari_2019BotSE} propose a chatbot, called Sayme, that resolves code conflicts among the software teams. Lin et al.~\cite{Lin_2020BotSE} propose MSA chatbot to assist practitioners in developing and maintaining the micro-services architecture. 

The increased usage of software chatbots in the software engineering domain motivates our work; to improve SE chatbot’s performance and help practitioners focus on the chatbot core functionalities rather than brainstorming more training queries. However, our work differs in that we propose an augmentation approach and do not develop a new chatbot.

\noindent\textbf{Datasets Augmentation.}
There is a number of studies that examine data augmentation techniques for text classification~\cite{Marivate_2020MLKE,feng_2020DeeLIO,Imran2022ASE,aminnejad_2020ACL,Rizos_2019CIKM,wei_2019IJCNLP}. For example, Marivate \& Sefara~\cite{Marivate_2020MLKE} evaluate the impact of four augmentation approaches (WordNet-based synonym, Word2vec-based, Round Trip Translation, and Mixup augmentation) on the performance of the classification algorithm using Sentiment 140, AG News, and Hate Speech datasets. Sharifirad et al.~\cite{sharifirad_2018ACL} propose an approach to improve the classification of sexiest tweets. In particular, the authors generate new tweets by replacing the words with their synonyms using the ConceptNet and Wikidata. The results show that applying the proposed approach improves the machine learning models in text classifications. Feng et al.~\cite{feng_2020DeeLIO} evaluate different augmentation approaches (e.g., random insertion) to fine-tune GPT-2 for text generation task using Yelp reviews dataset. The results show that keyword replacement and character-level synthetic noise are effective for text augmentation. Imran et. al~\cite{Imran2022ASE} evaluate the impact of data augmentation techniques (e.g., word insertion) to improve emotion classification. 
The work closest to ours is the work that augments datasets to enhance NLU's performance~\cite{malandrakis_2019ACL}. Malandrakis et al.~\cite{malandrakis_2019ACL} investigate the use of neural generative encoder-decoder models to improve the NLU's performance trained on movie and Live entertainment datasets. 

To the best of our knowledge, there is no work that studied evaluating the data augmentation approach that combines synonym replacement and paraphrasing techniques, and tailored it to improve the NLU's performance for SE chatbots. Our work differs and complements the prior work in two ways. First, our work imitates the chatbot developers by adding new synonyms and changing the sentence structure of the original training set. Second, we evaluate the approach using three SE datasets. Overall, our work complements the studies augmenting the training dataset to enhance NLU's performance by providing an augmentation approach for the SE domain.

\section{Threats to Validity}
\label{sec:threats6}
In this section, we discuss the threats to internal, replicability, and external validity of our study.

\noindent\textbf{Internal Validity.}
Concerns confounding factors that could have influenced our results. We configure BART to return three queries (sequences), which might impact the quality of the paraphrased queries. To alleviate this threat, the first two authors examined the output from BART using different numbers of returned queries (1, 3, 5, 7, and 10) and found that configuring the returned queries to be three yields the best results in terms sentence structure diversity and preserving the semantics. Another threat to internal validity relates to the manual labelling of our datasets, which could introduce subjectivity bias. However, these datasets have been used in many prior works in SE \cite{Farhour_FSE2024,Abdellatif_2021TSE}. Another potential threat is the choice of steps for our proposed approach and evaluation settings, which could impact the results. For example, using metrics other than Levenshtein distance (e.g., \cite{Bache2013KDD}) to measure the diversity of the candidate queries might impact the results. Therefore, we plan to evaluate using a wider range of settings (e.g., more scenarios) in the future.
In our study, we used three datasets (e.g., Quora Question Pairs) to fine-tune the BART transformer, as discussed in Section 4.3. Fine-tuning a transformer model involves some degree of randomness (e.g., random initialization of weights). Thus, replicating the study might lead to different results. We mitigate this issue by providing the scripts and datasets used in the evaluation in our replication package \cite{Augmenti93:online}. We believe that our study serves as a starting point for chatbot practitioners to leverage transformers in augmenting chatbot training dataset.

\noindent\textbf{External Validity}
Concerns the generalization of our findings. In this study, we evaluate an augmentation approach using three different datasets from the SE domain. The results might not generalize to other SE datasets. However, these datasets have been used by prior work to evaluate the performance of different NLUs and propose chatbots in the SE domain ~\cite{Abdellatif_2021TSE,Lin_2020BotSE,Dominic_2020BotSE}. Another threat is that we use Rasa NLU for evaluation; hence our results might not be generalizable to other NLUs. However, we select Rasa as it is an open-source NLU which guarantees that its internal implementation stays the same during our entire study. Moreover, Rasa has been widely used by practitioners to develop SE chatbots~\cite{Lin_2020BotSE,Abdellatif2020EMSE,Dominic_2020BotSE}.

\section{Conclusion \& Future Work}
\label{sec:conclusions6}
Software chatbots play important roles in the software engineering domain, enabling practitioners to perform various software development tasks, such as running tests, through natural language. The NLU component is essential for chatbots to understand the user's input. Training the NLU on possible queries from users is critical because it impacts user satisfaction with the chatbot. However, prior work shows that creating and augmenting a high-quality training dataset for SE chatbots is a costly and time-consuming task. To help chatbot developers improve the NLU's performance of their chatbots, we evaluate a transformer-based augmentation approach that emulates the standard way practitioners augment their chatbot training set. More specifically, the augmentation approach augments more queries by replacing words with synonyms (synonyms replacement) and paraphrasing the query using BART transformer. We evaluate the impact of using the augmentation approach on the NLU's performance using three datasets that represent distinct SE-related tasks. We find that the augmented queries provide a small but significant improvement in the NLU's performance. Moreover, the augmentation approach augments queries with diverse sentence structures while maintaining their original semantics (intents). Also, our results show that the augmentation approach increases the NLU's confidence in both correctly and incorrectly classified intents.

Our paper outlines directions for future work in this area. First, we plan to examine the performance of different transformers (e.g., GPT2, BERT, RoBERTa) in the task of paraphrasing SE queries. Also, we intend to investigate the use of Stack Overflow posts to craft a dataset containing paraphrased queries to help tune the transformers for the SE domain. Finally, we plan to evaluate the augmentation approach using more SE datasets and NLU platforms.

\bibliographystyle{ACM-Reference-Format}
\bibliography{bibliography}


\begin{thebibliography}{48}


\ifx \showCODEN    \undefined \def \showCODEN     #1{\unskip}     \fi
\ifx \showDOI      \undefined \def \showDOI       #1{#1}\fi
\ifx \showISBNx    \undefined \def \showISBNx     #1{\unskip}     \fi
\ifx \showISBNxiii \undefined \def \showISBNxiii  #1{\unskip}     \fi
\ifx \showISSN     \undefined \def \showISSN      #1{\unskip}     \fi
\ifx \showLCCN     \undefined \def \showLCCN      #1{\unskip}     \fi
\ifx \shownote     \undefined \def \shownote      #1{#1}          \fi
\ifx \showarticletitle \undefined \def \showarticletitle #1{#1}   \fi
\ifx \showURL      \undefined \def \showURL       {\relax}        \fi
\providecommand\bibfield[2]{#2}
\providecommand\bibinfo[2]{#2}
\providecommand\natexlab[1]{#1}
\providecommand\showeprint[2][]{arXiv:#2}

\bibitem[Abdellatif et~al\mbox{.}(2021)]%
        {Abdellatif_2021TSE}
\bibfield{author}{\bibinfo{person}{Ahmad Abdellatif}, \bibinfo{person}{Khaled Badran}, \bibinfo{person}{Diego Costa}, {and} \bibinfo{person}{Emad Shihab}.} \bibinfo{year}{2021}\natexlab{}.
\newblock \showarticletitle{A Comparison of Natural Language Understanding Platforms for Chatbots in Software Engineering}.
\newblock \bibinfo{journal}{\emph{IEEE Transactions on Software Engineering (TSE)}} (\bibinfo{year}{2021}), \bibinfo{pages}{1--1}.
\newblock
\urldef\tempurl%
\url{https://doi.org/10.1109/TSE.2021.3078384}
\showDOI{\tempurl}


\bibitem[Abdellatif et~al\mbox{.}(2020a)]%
        {Abdellatif2020EMSE}
\bibfield{author}{\bibinfo{person}{Ahmad Abdellatif}, \bibinfo{person}{Khaled Badran}, {and} \bibinfo{person}{Emad Shihab}.} \bibinfo{year}{2020}\natexlab{a}.
\newblock \showarticletitle{MSRBot: Using Bots to Answer Questions from Software Repositories}.
\newblock \bibinfo{journal}{\emph{Empirical Software Engineering (EMSE)}}  \bibinfo{volume}{25} (\bibinfo{year}{2020}), \bibinfo{pages}{1834--1863}.
\newblock
Issue 3.


\bibitem[Abdellatif et~al\mbox{.}(2020b)]%
        {Abdellatif_MSR2020}
\bibfield{author}{\bibinfo{person}{Ahmad Abdellatif}, \bibinfo{person}{Diego~Elias Costa}, \bibinfo{person}{Khaled Badran}, \bibinfo{person}{Rabe Abdelkareem}, {and} \bibinfo{person}{Emad Shihab}.} \bibinfo{year}{2020}\natexlab{b}.
\newblock \showarticletitle{Challenges in Chatbot Development: A Study of Stack Overflow Posts}. In \bibinfo{booktitle}{\emph{Proceedings of the 17th International Conference on Mining Software Repositories (MSR'20)}}. \bibinfo{pages}{To Appear}.
\newblock


\bibitem[Ahmad~Abdellatif(2024)]%
        {Augmenti93:online}
\bibfield{author}{\bibinfo{person}{Diego Costa Emad~Shihab Ahmad~Abdellatif, Khaled~Badran}.} \bibinfo{year}{2024}\natexlab{}.
\newblock \bibinfo{title}{A Transformer-based Approach for Augmenting Software Engineering Chatbots Datasets}.
\newblock \bibinfo{howpublished}{\url{https://zenodo.org/records/11121853}}.
\newblock
\newblock
\shownote{(Accessed on 05/06/2024)}.


\bibitem[Amin-Nejad et~al\mbox{.}(2020)]%
        {aminnejad_2020ACL}
\bibfield{author}{\bibinfo{person}{Ali Amin-Nejad}, \bibinfo{person}{Julia Ive}, {and} \bibinfo{person}{Sumithra Velupillai}.} \bibinfo{year}{2020}\natexlab{}.
\newblock \showarticletitle{Exploring Transformer Text Generation for Medical Dataset Augmentation}. In \bibinfo{booktitle}{\emph{Proceedings of the 12th Language Resources and Evaluation Conference}}. \bibinfo{publisher}{European Language Resources Association}, \bibinfo{address}{Marseille, France}, \bibinfo{pages}{4699--4708}.
\newblock
\showISBNx{979-10-95546-34-4}


\bibitem[Bache et~al\mbox{.}(2013)]%
        {Bache2013KDD}
\bibfield{author}{\bibinfo{person}{Kevin Bache}, \bibinfo{person}{David Newman}, {and} \bibinfo{person}{Padhraic Smyth}.} \bibinfo{year}{2013}\natexlab{}.
\newblock \showarticletitle{Text-based measures of document diversity}. In \bibinfo{booktitle}{\emph{Proceedings of the 19th ACM SIGKDD International Conference on Knowledge Discovery and Data Mining}} (Chicago, Illinois, USA) \emph{(\bibinfo{series}{KDD '13})}. \bibinfo{publisher}{Association for Computing Machinery}, \bibinfo{address}{New York, NY, USA}, \bibinfo{pages}{23–31}.
\newblock
\showISBNx{9781450321747}
\urldef\tempurl%
\url{https://doi.org/10.1145/2487575.2487672}
\showDOI{\tempurl}


\bibitem[Barash et~al\mbox{.}(2019)]%
        {Barash_2019FSE}
\bibfield{author}{\bibinfo{person}{Guy Barash}, \bibinfo{person}{Eitan Farchi}, \bibinfo{person}{Ilan Jayaraman}, \bibinfo{person}{Orna Raz}, \bibinfo{person}{Rachel Tzoref-Brill}, {and} \bibinfo{person}{Marcel Zalmanovici}.} \bibinfo{year}{2019}\natexlab{}.
\newblock \showarticletitle{Bridging the Gap between ML Solutions and Their Business Requirements Using Feature Interactions}. In \bibinfo{booktitle}{\emph{Proceedings of the 2019 27th ACM Joint Meeting on European Software Engineering Conference and Symposium on the Foundations of Software Engineering}} (Tallinn, Estonia) \emph{(\bibinfo{series}{ESEC/FSE 2019})}. \bibinfo{publisher}{Association for Computing Machinery}, \bibinfo{address}{New York, NY, USA}, \bibinfo{pages}{1048–1058}.
\newblock
\showISBNx{9781450355728}
\urldef\tempurl%
\url{https://doi.org/10.1145/3338906.3340442}
\showDOI{\tempurl}


\bibitem[Bora({[n.\,d.]})]%
        {geekprad94:online}
\bibfield{author}{\bibinfo{person}{Pradipta Bora}.} \bibinfo{year}{[n.\,d.]}\natexlab{}.
\newblock \bibinfo{title}{geekpradd/PyDictionary: PyDictionary is a Dictionary Module for Python 2/3 to get meanings, translations, synonyms and antonyms of words}.
\newblock \bibinfo{howpublished}{\url{https://github.com/geekpradd/PyDictionary}}.
\newblock
\newblock
\shownote{(Accessed on 07/01/2024)}.


\bibitem[Bradley et~al\mbox{.}(2018)]%
        {Bradley_2018ICSE}
\bibfield{author}{\bibinfo{person}{Nick~C. Bradley}, \bibinfo{person}{Thomas Fritz}, {and} \bibinfo{person}{Reid Holmes}.} \bibinfo{year}{2018}\natexlab{}.
\newblock \showarticletitle{Context-Aware Conversational Developer Assistants}. In \bibinfo{booktitle}{\emph{Proceedings of the 40th International Conference on Software Engineering}} (Gothenburg, Sweden) \emph{(\bibinfo{series}{ICSE '18})}. \bibinfo{publisher}{Association for Computing Machinery}, \bibinfo{address}{New York, NY, USA}, \bibinfo{pages}{993–1003}.
\newblock
\showISBNx{9781450356381}
\urldef\tempurl%
\url{https://doi.org/10.1145/3180155.3180238}
\showDOI{\tempurl}


\bibitem[Braun et~al\mbox{.}(2017)]%
        {braun_2017ACL}
\bibfield{author}{\bibinfo{person}{Daniel Braun}, \bibinfo{person}{Adrian Hernandez~Mendez}, \bibinfo{person}{Florian Matthes}, {and} \bibinfo{person}{Manfred Langen}.} \bibinfo{year}{2017}\natexlab{}.
\newblock \showarticletitle{Evaluating Natural Language Understanding Services for Conversational Question Answering Systems}. In \bibinfo{booktitle}{\emph{Proceedings of the 18th Annual {SIG}dial Meeting on Discourse and Dialogue}}. \bibinfo{publisher}{Association for Computational Linguistics}, \bibinfo{address}{Saarbr{\"u}cken, Germany}, \bibinfo{pages}{174--185}.
\newblock
\urldef\tempurl%
\url{https://doi.org/10.18653/v1/W17-5522}
\showDOI{\tempurl}


\bibitem[Ciniselli et~al\mbox{.}(2022a)]%
        {Ciniselli2022TSE}
\bibfield{author}{\bibinfo{person}{Matteo Ciniselli}, \bibinfo{person}{Nathan Cooper}, \bibinfo{person}{Luca Pascarella}, \bibinfo{person}{Antonio Mastropaolo}, \bibinfo{person}{Emad Aghajani}, \bibinfo{person}{Denys Poshyvanyk}, \bibinfo{person}{Massimiliano Di~Penta}, {and} \bibinfo{person}{Gabriele Bavota}.} \bibinfo{year}{2022}\natexlab{a}.
\newblock \showarticletitle{An Empirical Study on the Usage of Transformer Models for Code Completion}.
\newblock \bibinfo{journal}{\emph{IEEE Transactions on Software Engineering}} \bibinfo{volume}{48}, \bibinfo{number}{12} (\bibinfo{year}{2022}), \bibinfo{pages}{4818--4837}.
\newblock
\urldef\tempurl%
\url{https://doi.org/10.1109/TSE.2021.3128234}
\showDOI{\tempurl}


\bibitem[Ciniselli et~al\mbox{.}(2022b)]%
        {Ciniselli2022MSR}
\bibfield{author}{\bibinfo{person}{Matteo Ciniselli}, \bibinfo{person}{Luca Pascarella}, {and} \bibinfo{person}{Gabriele Bavota}.} \bibinfo{year}{2022}\natexlab{b}.
\newblock \showarticletitle{To What Extent Do Deep Learning-Based Code Recommenders Generate Predictions by Cloning Code from the Training Set?}. In \bibinfo{booktitle}{\emph{Proceedings of the 19th International Conference on Mining Software Repositories}} (Pittsburgh, Pennsylvania) \emph{(\bibinfo{series}{MSR '22})}. \bibinfo{publisher}{Association for Computing Machinery}, \bibinfo{address}{New York, NY, USA}, \bibinfo{pages}{167–178}.
\newblock
\showISBNx{9781450393034}
\urldef\tempurl%
\url{https://doi.org/10.1145/3524842.3528440}
\showDOI{\tempurl}


\bibitem[Docs(2022)]%
        {sentenceStructureLUIS_link}
\bibfield{author}{\bibinfo{person}{Microsoft Docs}.} \bibinfo{year}{2022}\natexlab{}.
\newblock \bibinfo{title}{Good example utterances}.
\newblock \bibinfo{howpublished}{\url{https://docs.microsoft.com/en-us/azure/cognitive-services/luis/luis-concept-utterance}}.
\newblock
\newblock
\shownote{(Accessed on 03/30/2024)}.


\bibitem[Dominic et~al\mbox{.}(2020a)]%
        {Dominic20BotSE}
\bibfield{author}{\bibinfo{person}{James Dominic}, \bibinfo{person}{Jada Houser}, \bibinfo{person}{Igor Steinmacher}, \bibinfo{person}{Charles Ritter}, {and} \bibinfo{person}{Paige Rodeghero}.} \bibinfo{year}{2020}\natexlab{a}.
\newblock \showarticletitle{Conversational Bot for Newcomers Onboarding to Open Source Projects}. In \bibinfo{booktitle}{\emph{Proceedings of the IEEE/ACM 42nd International Conference on Software Engineering Workshops}} (Seoul, Republic of Korea) \emph{(\bibinfo{series}{ICSEW'20})}. \bibinfo{publisher}{Association for Computing Machinery}, \bibinfo{address}{New York, NY, USA}, \bibinfo{pages}{46–50}.
\newblock
\showISBNx{9781450379632}
\urldef\tempurl%
\url{https://doi.org/10.1145/3387940.3391534}
\showDOI{\tempurl}


\bibitem[Dominic et~al\mbox{.}(2020b)]%
        {Dominic_2020BotSE}
\bibfield{author}{\bibinfo{person}{James Dominic}, \bibinfo{person}{Jada Houser}, \bibinfo{person}{Igor Steinmacher}, \bibinfo{person}{Charles Ritter}, {and} \bibinfo{person}{Paige Rodeghero}.} \bibinfo{year}{2020}\natexlab{b}.
\newblock \showarticletitle{Conversational Bot for Newcomers Onboarding to Open Source Projects}. In \bibinfo{booktitle}{\emph{Proceedings of the IEEE/ACM 42nd International Conference on Software Engineering Workshops}} (Seoul, Republic of Korea) \emph{(\bibinfo{series}{ICSEW'20})}. \bibinfo{publisher}{Association for Computing Machinery}, \bibinfo{address}{New York, NY, USA}, \bibinfo{pages}{46–50}.
\newblock
\showISBNx{9781450379632}
\urldef\tempurl%
\url{https://doi.org/10.1145/3387940.3391534}
\showDOI{\tempurl}


\bibitem[Dopierre et~al\mbox{.}(2021)]%
        {Dopierre_2021ACL}
\bibfield{author}{\bibinfo{person}{Thomas Dopierre}, \bibinfo{person}{C. Gravier}, {and} \bibinfo{person}{Wilfried Logerais}.} \bibinfo{year}{2021}\natexlab{}.
\newblock \showarticletitle{ProtAugment: Unsupervised diverse short-texts paraphrasing for intent detection meta-learning}. In \bibinfo{booktitle}{\emph{The 59th Annual Meeting of the Association for Computational Linguistics and the 11th International Joint Conference on Natural Language Processing}}, Vol.~\bibinfo{volume}{abs/2105.12995}.
\newblock


\bibitem[Efstathiou et~al\mbox{.}(2018)]%
        {Efstathiou_2018MSR}
\bibfield{author}{\bibinfo{person}{Vasiliki Efstathiou}, \bibinfo{person}{Christos Chatzilenas}, {and} \bibinfo{person}{Diomidis Spinellis}.} \bibinfo{year}{2018}\natexlab{}.
\newblock \showarticletitle{Word Embeddings for the Software Engineering Domain}. In \bibinfo{booktitle}{\emph{Proceedings of the 15th International Conference on Mining Software Repositories}} (Gothenburg, Sweden) \emph{(\bibinfo{series}{MSR '18})}. \bibinfo{publisher}{Association for Computing Machinery}, \bibinfo{address}{New York, NY, USA}, \bibinfo{pages}{38–41}.
\newblock
\showISBNx{9781450357166}
\urldef\tempurl%
\url{https://doi.org/10.1145/3196398.3196448}
\showDOI{\tempurl}


\bibitem[Farhour et~al\mbox{.}(2024)]%
        {Farhour_FSE2024}
\bibfield{author}{\bibinfo{person}{Farbod Farhour}, \bibinfo{person}{Ahmad Abdellatif}, \bibinfo{person}{Essam Mansour}, {and} \bibinfo{person}{Emad Shihab}.} \bibinfo{year}{2024}\natexlab{}.
\newblock \showarticletitle{A {Weak} {Supervision}-{Based} {Approach} to {Improve} {Chatbots} for {Code} {Repositories}}. In \bibinfo{booktitle}{\emph{Proceedings of the {ACM} {International} {Conference} on the {Foundations} of {Software} {Engineering} ({FSE}'24)}}.
\newblock


\bibitem[Fellbaum(1998)]%
        {wordnet_online}
\bibfield{author}{\bibinfo{person}{Christiane Fellbaum}.} \bibinfo{year}{1998}\natexlab{}.
\newblock \bibinfo{booktitle}{\emph{WordNet: An Electronic Lexical Database}}.
\newblock \bibinfo{publisher}{Bradford Books}.
\newblock
\urldef\tempurl%
\url{https://mitpress.mit.edu/9780262561167/}
\showURL{%
\tempurl}


\bibitem[Feng et~al\mbox{.}(2020)]%
        {feng_2020DeeLIO}
\bibfield{author}{\bibinfo{person}{Steven~Y. Feng}, \bibinfo{person}{Varun Gangal}, \bibinfo{person}{Dongyeop Kang}, \bibinfo{person}{Teruko Mitamura}, {and} \bibinfo{person}{Eduard Hovy}.} \bibinfo{year}{2020}\natexlab{}.
\newblock \showarticletitle{{G}en{A}ug: Data Augmentation for Finetuning Text Generators}. In \bibinfo{booktitle}{\emph{Proceedings of Deep Learning Inside Out (DeeLIO): The First Workshop on Knowledge Extraction and Integration for Deep Learning Architectures}}. \bibinfo{publisher}{Association for Computational Linguistics}, \bibinfo{address}{Online}, \bibinfo{pages}{29--42}.
\newblock
\urldef\tempurl%
\url{https://doi.org/10.18653/v1/2020.deelio-1.4}
\showDOI{\tempurl}


\bibitem[{Ilmania} et~al\mbox{.}(2018)]%
        {Ilmania_2018IALP}
\bibfield{author}{\bibinfo{person}{A. {Ilmania}}, \bibinfo{person}{{Abdurrahman}}, \bibinfo{person}{S. {Cahyawijaya}}, {and} \bibinfo{person}{A. {Purwarianti}}.} \bibinfo{year}{2018}\natexlab{}.
\newblock \showarticletitle{Aspect Detection and Sentiment Classification Using Deep Neural Network for Indonesian Aspect-Based Sentiment Analysis}. In \bibinfo{booktitle}{\emph{2018 International Conference on Asian Language Processing (IALP)}}. \bibinfo{pages}{62--67}.
\newblock
\urldef\tempurl%
\url{https://doi.org/10.1109/IALP.2018.8629181}
\showDOI{\tempurl}


\bibitem[Imran et~al\mbox{.}(2023)]%
        {Imran2022ASE}
\bibfield{author}{\bibinfo{person}{Mia~Mohammad Imran}, \bibinfo{person}{Yashasvi Jain}, \bibinfo{person}{Preetha Chatterjee}, {and} \bibinfo{person}{Kostadin Damevski}.} \bibinfo{year}{2023}\natexlab{}.
\newblock \showarticletitle{Data Augmentation for Improving Emotion Recognition in Software Engineering Communication}. In \bibinfo{booktitle}{\emph{Proceedings of the 37th IEEE/ACM International Conference on Automated Software Engineering}} (Rochester, MI, USA) \emph{(\bibinfo{series}{ASE '22})}. \bibinfo{publisher}{Association for Computing Machinery}, \bibinfo{address}{New York, NY, USA}, Article \bibinfo{articleno}{29}, \bibinfo{numpages}{13}~pages.
\newblock
\showISBNx{9781450394758}
\urldef\tempurl%
\url{https://doi.org/10.1145/3551349.3556925}
\showDOI{\tempurl}


\bibitem[jsphdnl(2023)]%
        {lackDataSOF_link1}
\bibfield{author}{\bibinfo{person}{jsphdnl}.} \bibinfo{year}{2023}\natexlab{}.
\newblock \bibinfo{title}{nlp Conversational Data for building a chat bot}.
\newblock \bibinfo{howpublished}{\url{https://stackoverflow.com/questions/45821517/conversational-data-for-building-a-chat-bot}}.
\newblock
\newblock
\shownote{(Accessed on 01/19/2024)}.


\bibitem[Larson et~al\mbox{.}(2019)]%
        {Larson_2019ACL}
\bibfield{author}{\bibinfo{person}{Stefan Larson}, \bibinfo{person}{Anish Mahendran}, \bibinfo{person}{Joseph~J. Peper}, \bibinfo{person}{Christopher Clarke}, \bibinfo{person}{Andrew Lee}, \bibinfo{person}{Parker Hill}, \bibinfo{person}{Jonathan~K. Kummerfeld}, \bibinfo{person}{Kevin Leach}, \bibinfo{person}{Michael~A. Laurenzano}, \bibinfo{person}{Lingjia Tang}, {and} \bibinfo{person}{Jason Mars}.} \bibinfo{year}{2019}\natexlab{}.
\newblock \showarticletitle{An Evaluation Dataset for Intent Classification and Out-of-Scope Prediction}. In \bibinfo{booktitle}{\emph{Proceedings of the 2019 Conference on Empirical Methods in Natural Language Processing and the 9th International Joint Conference on Natural Language Processing (EMNLP-IJCNLP)}}. \bibinfo{publisher}{Association for Computational Linguistics}, \bibinfo{address}{Hong Kong, China}, \bibinfo{pages}{1311--1316}.
\newblock
\urldef\tempurl%
\url{https://doi.org/10.18653/v1/D19-1131}
\showDOI{\tempurl}


\bibitem[Lebeuf et~al\mbox{.}(2018a)]%
        {Lebeuf2018IEEE}
\bibfield{author}{\bibinfo{person}{Carlene Lebeuf}, \bibinfo{person}{Margaret-Anne Storey}, {and} \bibinfo{person}{Alexey Zagalsky}.} \bibinfo{year}{2018}\natexlab{a}.
\newblock \showarticletitle{Software Bots}.
\newblock \bibinfo{journal}{\emph{IEEE Software}} \bibinfo{volume}{35}, \bibinfo{number}{1} (\bibinfo{year}{2018}), \bibinfo{pages}{18--23}.
\newblock
\urldef\tempurl%
\url{https://doi.org/10.1109/MS.2017.4541027}
\showDOI{\tempurl}


\bibitem[Lebeuf et~al\mbox{.}(2018b)]%
        {Lebeuf2018IEEESoftware}
\bibfield{author}{\bibinfo{person}{Carlene Lebeuf}, \bibinfo{person}{Margaret-Anne Storey}, {and} \bibinfo{person}{Alexey Zagalsky}.} \bibinfo{year}{2018}\natexlab{b}.
\newblock \showarticletitle{Software Bots}.
\newblock \bibinfo{journal}{\emph{IEEE Software}} \bibinfo{volume}{35}, \bibinfo{number}{1} (\bibinfo{year}{2018}), \bibinfo{pages}{18--23}.
\newblock
\urldef\tempurl%
\url{https://doi.org/10.1109/MS.2017.4541027}
\showDOI{\tempurl}


\bibitem[Lebeuf et~al\mbox{.}(2019)]%
        {Lebeuf2019BotSE}
\bibfield{author}{\bibinfo{person}{Carlene Lebeuf}, \bibinfo{person}{Alexey Zagalsky}, \bibinfo{person}{Matthieu Foucault}, {and} \bibinfo{person}{Margaret-Anne Storey}.} \bibinfo{year}{2019}\natexlab{}.
\newblock \showarticletitle{Defining and Classifying Software Bots: A Faceted Taxonomy}. In \bibinfo{booktitle}{\emph{Proceedings of the 1st International Workshop on Bots in Software Engineering}} (Montreal, Quebec, Canada) \emph{(\bibinfo{series}{BotSE '19})}. \bibinfo{publisher}{IEEE Press}, \bibinfo{pages}{1–6}.
\newblock
\urldef\tempurl%
\url{https://doi.org/10.1109/BotSE.2019.00008}
\showDOI{\tempurl}


\bibitem[Levenshtein(1966)]%
        {levenshtein1966SPD}
\bibfield{author}{\bibinfo{person}{Vladimir~I Levenshtein}.} \bibinfo{year}{1966}\natexlab{}.
\newblock \showarticletitle{Binary Codes Capable of Correcting Deletions, Insertions and Reversals}.
\newblock \bibinfo{journal}{\emph{Soviet Physics Doklady}}  \bibinfo{volume}{10} (\bibinfo{date}{February} \bibinfo{year}{1966}), \bibinfo{pages}{707}.
\newblock


\bibitem[Lewis et~al\mbox{.}(2020)]%
        {lewis_2020ACL}
\bibfield{author}{\bibinfo{person}{Mike Lewis}, \bibinfo{person}{Yinhan Liu}, \bibinfo{person}{Naman Goyal}, \bibinfo{person}{Marjan Ghazvininejad}, \bibinfo{person}{Abdelrahman Mohamed}, \bibinfo{person}{Omer Levy}, \bibinfo{person}{Veselin Stoyanov}, {and} \bibinfo{person}{Luke Zettlemoyer}.} \bibinfo{year}{2020}\natexlab{}.
\newblock \showarticletitle{{BART}: Denoising Sequence-to-Sequence Pre-training for Natural Language Generation, Translation, and Comprehension}. In \bibinfo{booktitle}{\emph{Proceedings of the 58th Annual Meeting of the Association for Computational Linguistics}}. \bibinfo{publisher}{Association for Computational Linguistics}, \bibinfo{address}{Online}, \bibinfo{pages}{7871--7880}.
\newblock
\urldef\tempurl%
\url{https://doi.org/10.18653/v1/2020.acl-main.703}
\showDOI{\tempurl}


\bibitem[Lin et~al\mbox{.}(2020)]%
        {Lin_2020BotSE}
\bibfield{author}{\bibinfo{person}{Chun-Ting Lin}, \bibinfo{person}{Shang-Pin Ma}, {and} \bibinfo{person}{Yu-Wen Huang}.} \bibinfo{year}{2020}\natexlab{}.
\newblock \showarticletitle{MSABot: A Chatbot Framework for Assisting in the Development and Operation of Microservice-Based Systems}. In \bibinfo{booktitle}{\emph{Proceedings of the IEEE/ACM 42nd International Conference on Software Engineering Workshops}} (Seoul, Republic of Korea) \emph{(\bibinfo{series}{BotSE'20})}. \bibinfo{publisher}{Association for Computing Machinery}, \bibinfo{address}{New York, NY, USA}, \bibinfo{pages}{36–40}.
\newblock
\showISBNx{9781450379632}
\urldef\tempurl%
\url{https://doi.org/10.1145/3387940.3391501}
\showDOI{\tempurl}


\bibitem[Malandrakis et~al\mbox{.}(2019)]%
        {malandrakis_2019ACL}
\bibfield{author}{\bibinfo{person}{Nikolaos Malandrakis}, \bibinfo{person}{Minmin Shen}, \bibinfo{person}{Anuj Goyal}, \bibinfo{person}{Shuyang Gao}, \bibinfo{person}{Abhishek Sethi}, {and} \bibinfo{person}{Angeliki Metallinou}.} \bibinfo{year}{2019}\natexlab{}.
\newblock \showarticletitle{Controlled Text Generation for Data Augmentation in Intelligent Artificial Agents}. In \bibinfo{booktitle}{\emph{Proceedings of the 3rd Workshop on Neural Generation and Translation}}. \bibinfo{publisher}{Association for Computational Linguistics}, \bibinfo{address}{Hong Kong}, \bibinfo{pages}{90--98}.
\newblock
\urldef\tempurl%
\url{https://doi.org/10.18653/v1/D19-5609}
\showDOI{\tempurl}


\bibitem[Marivate and Sefara(2020)]%
        {Marivate_2020MLKE}
\bibfield{author}{\bibinfo{person}{Vukosi Marivate} {and} \bibinfo{person}{Tshephisho Sefara}.} \bibinfo{year}{2020}\natexlab{}.
\newblock \showarticletitle{Improving Short Text Classification Through Global Augmentation Methods}. In \bibinfo{booktitle}{\emph{Machine Learning and Knowledge Extraction}}, \bibfield{editor}{\bibinfo{person}{Andreas Holzinger}, \bibinfo{person}{Peter Kieseberg}, \bibinfo{person}{A~Min Tjoa}, {and} \bibinfo{person}{Edgar Weippl}} (Eds.). \bibinfo{publisher}{Springer International Publishing}, \bibinfo{address}{Cham}, \bibinfo{pages}{385--399}.
\newblock
\showISBNx{978-3-030-57321-8}


\bibitem[Microsoft(2023)]%
        {Utteranc5:online}
\bibfield{author}{\bibinfo{person}{Microsoft}.} \bibinfo{year}{2023}\natexlab{}.
\newblock \bibinfo{title}{Utterances - Azure Cognitive Services | Microsoft Docs}.
\newblock \bibinfo{howpublished}{\url{https://docs.microsoft.com/en-us/azure/cognitive-services/luis/concepts/utterances}}.
\newblock
\newblock
\shownote{(Accessed on 01/08/2024)}.


\bibitem[Paikari et~al\mbox{.}(2019)]%
        {Paikari_2019BotSE}
\bibfield{author}{\bibinfo{person}{Elahe Paikari}, \bibinfo{person}{JaeEun Choi}, \bibinfo{person}{SeonKyu Kim}, \bibinfo{person}{Sooyoung Baek}, \bibinfo{person}{MyeongSoo Kim}, \bibinfo{person}{SeungEon Lee}, \bibinfo{person}{ChaeYeon Han}, \bibinfo{person}{YoungJae Kim}, \bibinfo{person}{KaHye Ahn}, \bibinfo{person}{Chan Cheong}, {and} \bibinfo{person}{André van~der hoek}.} \bibinfo{year}{2019}\natexlab{}.
\newblock \showarticletitle{A Chatbot for Conflict Detection and Resolution}. In \bibinfo{booktitle}{\emph{2019 IEEE/ACM 1st International Workshop on Bots in Software Engineering (BotSE)}}. \bibinfo{pages}{29--33}.
\newblock
\urldef\tempurl%
\url{https://doi.org/10.1109/BotSE.2019.00016}
\showDOI{\tempurl}


\bibitem[Rasa({[n.\,d.]})]%
        {Conversa24:online}
\bibfield{author}{\bibinfo{person}{Rasa}.} \bibinfo{year}{[n.\,d.]}\natexlab{}.
\newblock \bibinfo{title}{Conversational AI Platform | Superior Customer Experiences Start Here}.
\newblock \bibinfo{howpublished}{\url{https://rasa.com/}}.
\newblock
\newblock
\shownote{(Accessed on 07/01/2024)}.


\bibitem[Rizos et~al\mbox{.}(2019)]%
        {Rizos_2019CIKM}
\bibfield{author}{\bibinfo{person}{Georgios Rizos}, \bibinfo{person}{Konstantin Hemker}, {and} \bibinfo{person}{Bj\"{o}rn Schuller}.} \bibinfo{year}{2019}\natexlab{}.
\newblock \showarticletitle{Augment to Prevent: Short-Text Data Augmentation in Deep Learning for Hate-Speech Classification}. In \bibinfo{booktitle}{\emph{Proceedings of the 28th ACM International Conference on Information and Knowledge Management}} (Beijing, China) \emph{(\bibinfo{series}{CIKM '19})}. \bibinfo{publisher}{Association for Computing Machinery}, \bibinfo{address}{New York, NY, USA}, \bibinfo{pages}{991–1000}.
\newblock
\showISBNx{9781450369763}
\urldef\tempurl%
\url{https://doi.org/10.1145/3357384.3358040}
\showDOI{\tempurl}


\bibitem[Sharifirad et~al\mbox{.}(2018)]%
        {sharifirad_2018ACL}
\bibfield{author}{\bibinfo{person}{Sima Sharifirad}, \bibinfo{person}{Borna Jafarpour}, {and} \bibinfo{person}{Stan Matwin}.} \bibinfo{year}{2018}\natexlab{}.
\newblock \showarticletitle{Boosting Text Classification Performance on Sexist Tweets by Text Augmentation and Text Generation Using a Combination of Knowledge Graphs}. In \bibinfo{booktitle}{\emph{Proceedings of the 2nd Workshop on Abusive Language Online ({ALW}2)}}. \bibinfo{publisher}{Association for Computational Linguistics}, \bibinfo{address}{Brussels, Belgium}, \bibinfo{pages}{107--114}.
\newblock
\urldef\tempurl%
\url{https://doi.org/10.18653/v1/W18-5114}
\showDOI{\tempurl}


\bibitem[Sheri(2022)]%
        {lackDataSOF_link3}
\bibfield{author}{\bibinfo{person}{Sheri}.} \bibinfo{year}{2022}\natexlab{}.
\newblock \bibinfo{title}{Python Intent classification for Chatbot}.
\newblock \bibinfo{howpublished}{\url{https://stackoverflow.com/questions/62970861/intent-classification-for-chatbot}}.
\newblock
\newblock
\shownote{(Accessed on 04/28/2024)}.


\bibitem[Shridhar et~al\mbox{.}(2020)]%
        {shridhar_2020ACL}
\bibfield{author}{\bibinfo{person}{Kumar Shridhar}, \bibinfo{person}{Harshil Jain}, \bibinfo{person}{Akshat Agarwal}, {and} \bibinfo{person}{Denis Kleyko}.} \bibinfo{year}{2020}\natexlab{}.
\newblock \showarticletitle{End to End Binarized Neural Networks for Text Classification}. In \bibinfo{booktitle}{\emph{Proceedings of SustaiNLP: Workshop on Simple and Efficient Natural Language Processing}}. \bibinfo{publisher}{Association for Computational Linguistics}, \bibinfo{address}{Online}, \bibinfo{pages}{29--34}.
\newblock
\urldef\tempurl%
\url{https://doi.org/10.18653/v1/2020.sustainlp-1.4}
\showDOI{\tempurl}


\bibitem[Storey and Zagalsky(2016)]%
        {Storey2016FSE}
\bibfield{author}{\bibinfo{person}{Margaret-Anne Storey} {and} \bibinfo{person}{Alexey Zagalsky}.} \bibinfo{year}{2016}\natexlab{}.
\newblock \showarticletitle{Disrupting Developer Productivity One Bot at a Time}. In \bibinfo{booktitle}{\emph{Proceedings of the 2016 24th ACM SIGSOFT International Symposium on Foundations of Software Engineering}} (Seattle, WA, USA) \emph{(\bibinfo{series}{FSE 2016})}. \bibinfo{publisher}{Association for Computing Machinery}, \bibinfo{address}{New York, NY, USA}, \bibinfo{pages}{928–931}.
\newblock
\showISBNx{9781450342186}
\urldef\tempurl%
\url{https://doi.org/10.1145/2950290.2983989}
\showDOI{\tempurl}


\bibitem[Tmbo(2022)]%
        {sentenceStructureRasa_link}
\bibfield{author}{\bibinfo{person}{Tmbo}.} \bibinfo{year}{2022}\natexlab{}.
\newblock \bibinfo{title}{multiple entity recognition· Issue \#427 · RasaHQ/rasa}.
\newblock \bibinfo{howpublished}{\url{https://github.com/RasaHQ/rasa/issues/427}}.
\newblock
\newblock
\shownote{(Accessed on 03/23/2024)}.


\bibitem[Tufano et~al\mbox{.}(2022)]%
        {Tufano2022ICSE}
\bibfield{author}{\bibinfo{person}{Rosalia Tufano}, \bibinfo{person}{Simone Masiero}, \bibinfo{person}{Antonio Mastropaolo}, \bibinfo{person}{Luca Pascarella}, \bibinfo{person}{Denys Poshyvanyk}, {and} \bibinfo{person}{Gabriele Bavota}.} \bibinfo{year}{2022}\natexlab{}.
\newblock \showarticletitle{Using Pre-Trained Models to Boost Code Review Automation}. In \bibinfo{booktitle}{\emph{Proceedings of the 44th International Conference on Software Engineering}} (Pittsburgh, Pennsylvania) \emph{(\bibinfo{series}{ICSE '22})}. \bibinfo{publisher}{Association for Computing Machinery}, \bibinfo{address}{New York, NY, USA}, \bibinfo{pages}{2291–2302}.
\newblock
\showISBNx{9781450392211}
\urldef\tempurl%
\url{https://doi.org/10.1145/3510003.3510621}
\showDOI{\tempurl}


\bibitem[Wei and Zou(2019)]%
        {wei_2019IJCNLP}
\bibfield{author}{\bibinfo{person}{Jason Wei} {and} \bibinfo{person}{Kai Zou}.} \bibinfo{year}{2019}\natexlab{}.
\newblock \showarticletitle{{EDA}: Easy Data Augmentation Techniques for Boosting Performance on Text Classification Tasks}. In \bibinfo{booktitle}{\emph{Proceedings of the 2019 Conference on Empirical Methods in Natural Language Processing and the 9th International Joint Conference on Natural Language Processing (EMNLP-IJCNLP)}}. \bibinfo{publisher}{Association for Computational Linguistics}, \bibinfo{address}{Hong Kong, China}, \bibinfo{pages}{6382--6388}.
\newblock
\urldef\tempurl%
\url{https://doi.org/10.18653/v1/D19-1670}
\showDOI{\tempurl}


\bibitem[West et~al\mbox{.}(2021)]%
        {West_2021ACL}
\bibfield{author}{\bibinfo{person}{Peter West}, \bibinfo{person}{Ximing Lu}, \bibinfo{person}{Ari Holtzman}, \bibinfo{person}{Chandra Bhagavatula}, \bibinfo{person}{Jena~D. Hwang}, {and} \bibinfo{person}{Yejin Choi}.} \bibinfo{year}{2021}\natexlab{}.
\newblock \showarticletitle{Reflective Decoding: Beyond Unidirectional Generation with Off-the-Shelf Language Models}. In \bibinfo{booktitle}{\emph{Proceedings of the 59th Annual Meeting of the Association for Computational Linguistics and the 11th International Joint Conference on Natural Language Processing (Volume 1: Long Papers)}}. \bibinfo{publisher}{Association for Computational Linguistics}, \bibinfo{address}{Online}, \bibinfo{pages}{1435--1450}.
\newblock
\urldef\tempurl%
\url{https://doi.org/10.18653/v1/2021.acl-long.114}
\showDOI{\tempurl}


\bibitem[Wilks(2011)]%
        {wilks2011statistical}
\bibfield{author}{\bibinfo{person}{Daniel~S Wilks}.} \bibinfo{year}{2011}\natexlab{}.
\newblock \bibinfo{booktitle}{\emph{Statistical methods in the atmospheric sciences}}. Vol.~\bibinfo{volume}{100}.
\newblock \bibinfo{publisher}{Academic press}.
\newblock


\bibitem[Ye et~al\mbox{.}(2016)]%
        {Ye_2016SANER}
\bibfield{author}{\bibinfo{person}{Deheng Ye}, \bibinfo{person}{Zhenchang Xing}, \bibinfo{person}{Chee~Yong Foo}, \bibinfo{person}{Zi~Qun Ang}, \bibinfo{person}{Jing Li}, {and} \bibinfo{person}{Nachiket Kapre}.} \bibinfo{year}{2016}\natexlab{}.
\newblock \showarticletitle{Software-Specific Named Entity Recognition in Software Engineering Social Content}. In \bibinfo{booktitle}{\emph{2016 IEEE 23rd International Conference on Software Analysis, Evolution, and Reengineering (SANER)}}, Vol.~\bibinfo{volume}{1}. \bibinfo{pages}{90--101}.
\newblock
\urldef\tempurl%
\url{https://doi.org/10.1109/SANER.2016.10}
\showDOI{\tempurl}


\bibitem[Zhang et~al\mbox{.}(2019)]%
        {Zhang_2019naacl}
\bibfield{author}{\bibinfo{person}{Yuan Zhang}, \bibinfo{person}{Jason Baldridge}, {and} \bibinfo{person}{Luheng He}.} \bibinfo{year}{2019}\natexlab{}.
\newblock \showarticletitle{{PAWS: Paraphrase Adversaries from Word Scrambling}}. In \bibinfo{booktitle}{\emph{Proc. of NAACL}}.
\newblock


\bibitem[Zhou et~al\mbox{.}(2020)]%
        {Zhou_2020ACL}
\bibfield{author}{\bibinfo{person}{Jianing Zhou}, \bibinfo{person}{Hongyu Gong}, {and} \bibinfo{person}{Suma Bhat}.} \bibinfo{year}{2020}\natexlab{}.
\newblock \showarticletitle{PIE: A Parallel Idiomatic Expression Corpus for Idiomatic Sentence Generation and Paraphrasing}. In \bibinfo{booktitle}{\emph{The Joint Conference of the 59th Annual Meeting of the Association for Computational Linguistics and the 11th International Joint Conference on Natural Language Processing (ACL-IJCNLP 2021), MWE Workshop, 2021}}.
\newblock


\end{thebibliography}

\end{document}